%% file: aanda.tex
\documentclass{aa}  
\usepackage{graphicx}
\usepackage{txfonts}
\newcommand{\teff}{\ensuremath{\mathrm{T_{eff}}}}
\newcommand{\kms}{\ensuremath{\mathrm{km\,s^{-1}}}}
\newcommand{\logg}{\ensuremath{\mathrm{\log g}}}
\newcommand{\vt}{\ensuremath{\mathrm{v_{turb}}}}

\usepackage{xcolor}

\begin{document}

   \title{Tracing Pop~III supernovae with extreme energies \\ through the Sculptor dwarf spheroidal galaxy\thanks{ESO ID 107.22SJ; Based on observations made with ESO VLT at the La Silla Paranal observatory under programs ID 107.22SJ, and  
   0101.B-0189, 0101.D-0210, as well as on data obtained from the ESO Science Archive Facility}}

\titlerunning{The early chemical enrichment of Sculptor}
%   \subtitle{I. Overviewing the $\kappa$-mechanism}

   \author{\'{A}. Sk\'{u}lad\'{o}ttir
          \inst{1,2}
          \and
          I. Vanni\inst{1,2}
          \and
          S. Salvadori\inst{1,2}
          \and
          R. Lucchesi\inst{1}
          }

   \institute{
    Dipartimento di Fisica e Astronomia, Universit\'a degli Studi di Firenze, Via G. Sansone 1, I-50019 Sesto Fiorentino, Italy. \\ \email{asa.skuladottir@unifi.it}
    \and
    INAF/Osservatorio Astrofisico di Arcetri, Largo E. Fermi 5, I-50125 Firenze, Italy.\\
             }

%   \date{Received September 15, 1996; accepted March 16, 1997}

% \abstract{}{}{}{}{} 
% 5 {} token are mandatory
 
  \abstract{
 The Sculptor dwarf spheroidal galaxy is old and metal-poor, making it ideal to study the earliest chemical enrichment in the Local Group. We followed up the most metal-poor star known in this (or any external) galaxy, AS0039, with high-resolution ESO VLT/UVES spectra. Our new analysis confirmed its low metallicity, $\rm[Fe/H]_{LTE}=-3.90\pm0.15$, and that it is extremely C-poor, with $\rm A(C)=+3.60$, which corresponds to $\rm[C/Fe]_{LTE}=-0.33\pm0.17$ (accounting for internal mixing). This adds to the evidence of Sculptor being intrinsically C-poor at low $\rm[Fe/H]\lesssim-3$. However, here we also report a new discovery of a carbon-enhanced metal-poor star in Sculptor, DR20080, with no enhancement of Ba (CEMP-no), indicative of enrichment by zero-metallicity low-energy supernovae, $E_\textsl{SN} < 1 \times 10^{51}$. This is the first evidence of a dual population of CEMP-no and C-normal stars in Sculptor at $\rm[Fe/H]\leq{-3}$. The fraction of CEMP-no stars is still low, $f^\textsl{Scl}_\textsl{CEMP}=9^{+11}_{-8}\%$ at $\rm -4\leq[Fe/H]\leq-3$, compared to the significantly higher fraction in the Milky Way halo, $f^\textsl{MW}_\textsl{CEMP}\approx40\%$. To further investigate the early chemical enrichment of Sculptor, we re-derive chemical abundances of light, $\alpha$-, iron peak, and neutron-capture elements in all Sculptor stars at $\rm [Fe/H]\leq-2.8$, with available high-resolution spectra. Our results show that at these low [Fe/H], Sculptor is deficient in light elements (e.g. C, Na, Al, Mg) relative to both the Milky Way halo, and ultra-faint dwarf galaxies, pointing towards significant contribution of high-energy supernovae. Furthermore, the abundance pattern of the star AS0039 is best fitted with a zero-metallicity hypernova progenitor, $E_\textsl{SN} = 10 \times 10^{51}$, with a mass of $M=20$\,M$_\odot$. Our results in Sculptor, at $\rm[Fe/H]\leq-3$, therefore suggest significant enrichment by both very low-energy supernovae and hypernovae, solidifying this galaxy as one of the benchmarks for understanding the energy distribution of the first supernova in the Universe.
 
  }


   \maketitle
%
%-------------------------------------------------------------------

\section{Introduction}
The properties of the first stars in the Universe remain elusive. Although the impact of the first stellar generation was significant, as they provided the first ionizing radiation, metals and dust in the Universe, their study is challenging. No metal-free (Pop~III) star has been observed to date, and it is unclear whether the newly launched JWST telescope will be able to directly observe Pop~III galaxies \citep[][]{Wang2023} or Pop~III supernovae (SN; e.g. \citealt{Regos20, Yan2023}).

At the present moment, the most convincing observational constraints for the properties of Pop~III stars come from observing ancient stars in the Milky Way and its satellite galaxies. In particular, the carbon-enhanced metal-poor stars (CEMP-no; $\rm[C/Fe]>+0.7$, no Ba or Eu enhancements; e.g. \citealt{BeersChristlieb05,Norris13,Bonifacio2015,NorrisYong19}) are commonly accepted as the direct descendants of Pop~III stars, $10\lesssim M_{\star}/M_\odot\lesssim100$, which ended their lives as faint supernovae, polluting their environments mainly with C and the lighter elements ($Z<20$) resulting in very high [C/Fe] ratios \citep[e.g.][]{Iwamoto2005a,Marassi2014a,DeBennassuti2017,Hartwig2018,Welsh2022,Rossi2023}. The fraction of these CEMP-no stars in the Milky Way becomes higher towards lower metallicity, with $f^\textsl{MW}_\textsl{CEMP}=40\%$ at $\rm [Fe/H]\leq-3$, and as high as $f^\textsl{MW}_\textsl{CEMP}=80\%$ at $\rm [Fe/H]\leq-4$ \citep[e.g.][]{Placco14}. The small ultra-faint dwarf galaxies (UFD) seem to follow the same pattern, with very high ratios of CEMP-no stars, compatible with the Milky Way halo (e.g.~\citealt{Salvadori15,Ji20}). However, the situation is less clear in the more massive dwarf galaxies.

\begin{figure*}[ht]
\centering
\includegraphics[width=0.45\linewidth]{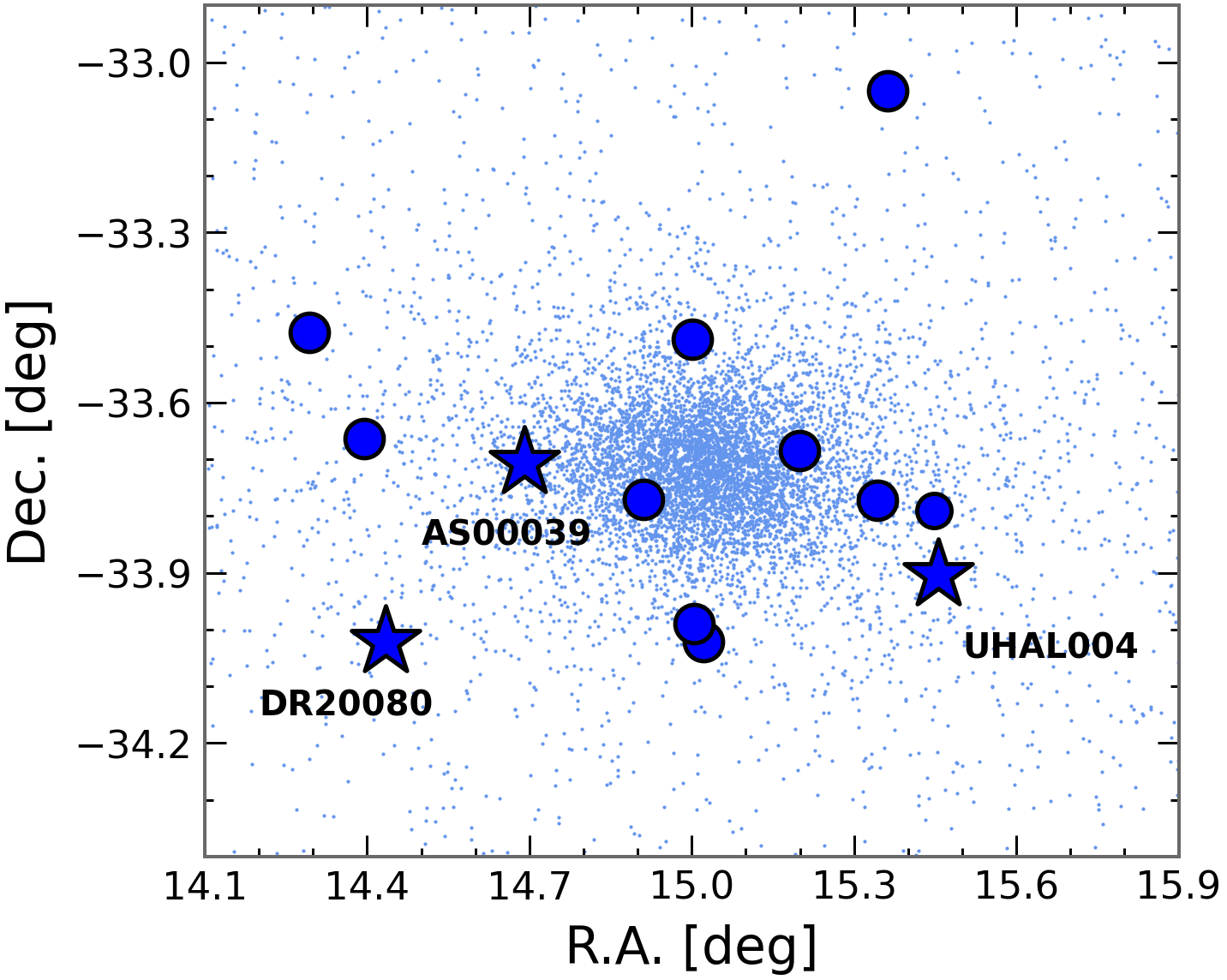}
\includegraphics[width=0.45\linewidth]{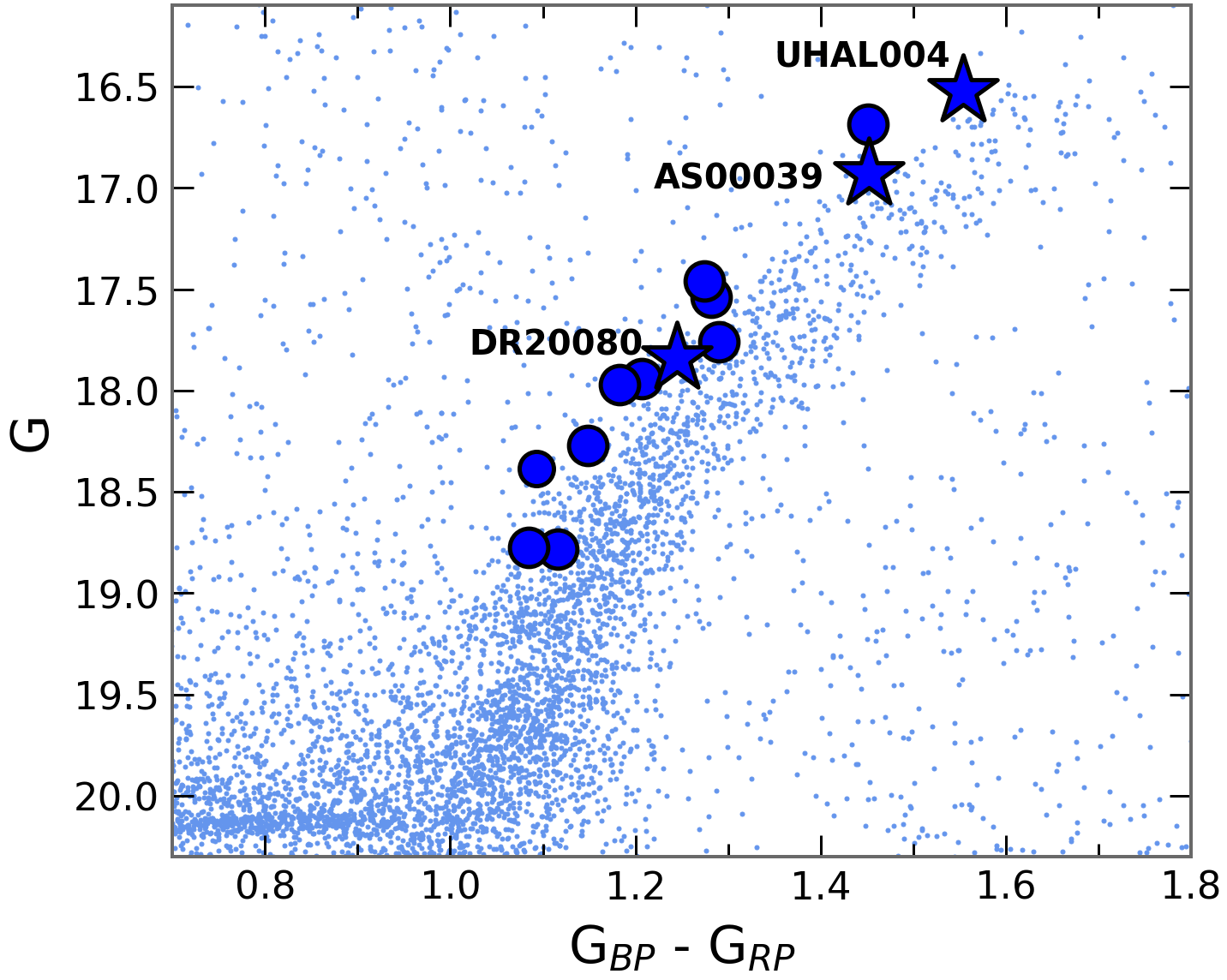}
\caption{Position on the sky (\textit{left}) and location on the RGB (\textit{right}) for our Sculptor stellar sample, based on \textit{Gaia} DR3 (Table~\ref{tab:stellsample}). Star symbols with labels are stars with new spectra, for the first time analysed here; while large blue circles are stars with available literature spectra, reanalysed here. Small, light blue points are stars from \textit{Gaia} eDR3 within 1.5 deg radius of the center of Sculptor.}
%from metov.py
\label{fig:radecrgb}
\end{figure*}

The Sculptor dwarf spheroidal (dSph) galaxy is one of the the best studied systems at low metallicity \citep{Tafelmeyer10,Frebel10,KirbyCohen12,Starkenburg13,Skuladottir15a,Skuladottir21,Jablonka15,Simon15,Chiti18}. Its stellar population is predominantly old, $>10\,$ Gyr \citep{deBoer12,Bettinelli19}, metal-poor, $\rm \langle[Fe/H]=-1.8\rangle$, and its chemical abundances are relatively homogeneous \citep{Kirby09,Kirby10,Skuladottir15b,Skuladottir17,Skuladottir18,Skuladottir19,Hill19,delosReyes20}. Until now, however, only one very metal-poor ($\rm [Fe/H]\leq-2$) star has been discovered to be C-rich relative to the normal population (and with $\rm[Ba/Fe]<0$), i.e. the CEMP-no star ET0097 at $\rm [Fe/H]=-2$ \citep{Skuladottir15a}, while none has been confirmed with intermediate- to high-resolution spectroscopy at $\rm[Fe/H]\leq-3$. The fraction of CEMP-no stars therefore seems significantly smaller than in the Milky Way, $f^\textsl{Scl}_\textsl{CEMP}<25\%$ at $\rm [Fe/H]\leq-3$ \citep{Starkenburg13,Skuladottir15a}. 

As valuable as the CEMP-no stars are, they are only expected to probe a limited fraction of the parameter space of the first stars, characterised by low energy of their supernovae. The Pop~III stars are predicted to end their lives as SN of a range of explosion energies \citep{HegerWoosley02}, but this remains poorly explored. However, recently, stars have been identified in Sculptor \citep{Skuladottir21} and the Milky Way halo \citep{Placco21,Mardini22} which show evidence of being imprinted by high-energy SNe. It still remains unclear whether the smaller UFDs would be able to retain the products of such high energy Pop~III SNe (e.g.~Rossi et al. in prep.).
The initial mass function (IMF) at birth of Pop~III stars has been predicted to be top-heavy \citep[e.g.][]{Hirano15,Sharda22,Sharda23}, i.e. that the first stars were typically more massive than present day stars. This conclusion is supported by models using the non-detection of Pop~III to constrain the IMF of the first stars \citep{Magg18,Magg19,Rossi21}. In particular, Pop~III stars in the mass range $140\leq M_\star/M_\odot \leq 260$ are predicted to explode as pair-instability supernovae (PISN), enriching their environment with unique yields, showing very strong odd-even effect \citep[e.g.][]{HegerWoosley02,Salvadori19,Aguado2023b}. Small UFD galaxies are unlikely to be able to retain the products of the very energetic PISN. Their descendants are therefore more likely to be found in more massive systems such as dSph galaxies, or the Galactic bulge \citep{Pagnini23}.
%The Milky Way halo is believed to be built up from accreted galaxies, covering a large range of masses \citep[e.g.][]{Deason16}, and thus descendants of all types of Pop~III stars should be present.}
%In a hierarchical \Lambda CDM Universe UFDs are predicted to be the progenitors of more massive galaxies, including classical dSphs \citep[e.g.][]{Salvadori15}, which are predicted by some models to be the main contributors of the stellar halo \citep[e.g.][]{Deason16}. 

%Small UFD galaxies are unlikely to be able to retain the products of the very energetic PISN, and the descendants are more likely to be found in systems such as dSph galaxies, or the Galactic bulge \citep[see][]{Pagnini23}. \red{\bf The Milky Way halo is likely built up from accreted galaxies, covering a large range of masses, and thus descendants of all types of Pop~III stars should be present \citep[e.g.][]{Salvadori19}. The dominant contributors to the halo are predicted to be dwarf galaxies more massive than Sculptor, i.e. $M_\star\sim10^8- 10^{10}$\,M$_\odot$, and UFDs are expected to contribute $<5$\%, even at $\rm[Fe/H]<-2$ \citep{Deason16}. The comparable CEMP-no fraction in UFDs and the Milky Way halo could possibly contradict this picture, however, observations in more massive dwarf galaxies, at $\rm [Fe/H]<-3$, are still severely lacking.}

The ancient and metal-poor Sculptor dSph offers a unique opportunity to expand our current knowledge of the first population of stars in the Universe. Being significantly larger, $M_{tot}\gtrsim 4\times 10^8$ \citep{Battaglia08} than the UFD galaxies, Sculptor is likely to retain more of the products of high-energy SN, which might be lost in galaxies with a smaller potential well. Currently, Sculptor is the external galaxy with most known stars at $\rm [Fe/H]\leq-3$, many of which have been observed with high-resolution spectra, necessary for detailed and high-precision abundance analysis. Here we add pieces to the Pop~III puzzle with: (i) an analysis of a new high-quality spectra of one of the most metal-poor star in any external galaxy, AS0039; (ii) the discovery of the first CEMP-no star in Sculptor at $\rm[Fe/H]\leq-3$; and (iii) a homogeneous (re)analysis of the 11 most metal-poor stars in Sculptor which have high-resolution ($R \gtrsim 20\,000$) spectra, including one unpublished star at $\rm[Fe/H]=-2.8$. Thereby we provide the most complete picture of the earliest chemical evolution of any external galaxy to date.

\input tables/newdata.tex
\input tables/stellsample.tex

%--------------------------------------------------------------------

\section{Observational data}

\subsection{New data}

Here we present previously unpublished data of three Sculptor stars: AS0039, DR20080 and UHAL004. Their spectra is described in Table~\ref{tab:newdata}, while their basic properties, such as positions and photometric values are in Table~\ref{tab:stellsample} and Fig.~\ref{fig:radecrgb}, according to \textit{Gaia} Early Data Release 3 (eDR3; \citealt{eDR3}). 

The stars AS00039 and DR20080 in the Sculptor dSph were discovered to be metal-poor through a large ESO VLT/FLAMES survey of the \ion{Ca}{II} near-infrared triplet (ESO ID 0102.B-0786; \citealt{Tolstoy23}). They were followed up by the same programme with VLT/X-Shooter, for confirmation of their low-metallicity, and here we report basic results (Fe~and~C) for DR20080 using this spectrum. The discovery X-Shooter spectrum of AS0039 was analysed and the results published in \citet{Skuladottir21}. Here we present an analysis based on a new, higher quality UVES spectrum. The star AS00039 shows evidence of being a binary, with $v_\textit{rad}=130.5\pm1.0\,\kms$ (Aug-Sep 2021), significantly different from $v_\textit{rad}=135\pm1\,\kms$ reported from the X-Shooter spectrum (Dec 2018; \citealt{Skuladottir21}). Gaia identifies one other star within 5'' from AS0039, however its colours are not compatible with the RGB of the Sculptor dSph ($G=19.02$, $G_{BP}-G_{RP}=1.96$; see Fig.~\ref{fig:radecrgb}), so the companion star is not identified. Finally, the star UHAL004 was observed via the VLT/FLAMES program with ID 0101.D-0210(A), and discovered to be very metal-poor. The available spectra consists of FLAMES/UVES and FLAMES/GIRAFFE spectra, as listed in Table~\ref{tab:newdata}.

\subsection{Archival data}

To have a clear and homogeneous picture of the earliest chemical enrichment in Sculptor, spectra of all previously published stars with $\rm [Fe/H]\leq-3$ \citep{Tafelmeyer10,Starkenburg13,Jablonka15} were retrieved from the ESO archive. In addition, fully reduced spectra for three stars, observed with Magellan/MIKE, were obtained through private communication w/ A. Frebel and J. Simon \citep{Frebel10,Simon15}. For convenience, all archival stars are given new names here, referring to the initials of the first author where the spectra we use was first presented (see Table~\ref{tab:stellsample}). 

Several of the archival stars have been analysed in more than one publication, either with the same spectra \citep{Tafelmeyer10,Frebel10,Simon15}, or different spectra \citep{Starkenburg13,Jablonka15}. Details of the spectral properties are listed in the respective papers. We note that the star ES03170 does not, to our knowledge, have available high resolution ($R\gtrsim20\,000$) spectra, and the star MT00750 is lacking spectral coverage to measure the C abundance. The positions and magnitudes of the entire stellar sample is listed in Table~\ref{tab:stellsample}. Finally, we note that the stars JS14296 and JS66402 are the faintest of the HR sample, and accordingly also those with the lowest S/N \citep{Simon15}.

\input tables/abundances.tex

\section{Atmospheric Parameters} \label{sec:stellpar}

\subsection{Effective temperature \teff}

The effective temperatures, \teff, were determined using photometry from Gaia eDR3 \citep{eDR3} and the $G_{BP}-G_{RP}$ calibration from \citet{MucciarelliBellazzini20}. We evaluate the extinction by converting the known extinction of $E(V-I)=0.027$ in the direction of Sculptor\footnote{\url{https://ned.ipac.caltech.edu/forms/calculator.html}} \citep{Schlafly11} to $E(G_{BP}-G_{RP})=0.024$ \citep{GaiaBusso18}. A typical random error of $\Delta\teff=86$\,K is adopted, as the quadratic sum arising from: uncertainties in the the photometry, 0.01 mag for $G_{BP}$ and $G_{RP}$ \citep{eDR3}; an error on [Fe/H], 0.20 dex; and the $\sigma_\teff$ of the $G_{BP}-G_{RP}$ calibration for giants, 83\,K \citep{MucciarelliBellazzini20}.

Spectroscopic temperatures have been shown to be typically lower than the photometric scales. Furthermore, \citet{MucciarelliBonifacio20} showed that this difference increases towards lower metallicity, and concluded that this is likely a direct consequence of 3D NLTE effects of individual \ion{Fe}{I} lines. This effect can be limited by rejecting lines with excitation potential $<1.4$\,eV, as done e.g. in \citet{Tafelmeyer10} and \citet{Jablonka15}. As expected, our photometric \teff\ scale is thus higher compared to those that used the spectroscopic method, $\langle\Delta\teff\rangle \,\approx+200$\,K 
 compared to the works of \citet{Frebel10}, \citet{Tafelmeyer10}, and \citet{Simon15}; and $\langle\Delta\teff\rangle \,\approx+100$\,K with \citet{Jablonka15}; while our \teff\ scale agreed on average within 50\,K with the photometric \teff\ of \citet{Starkenburg13}.

\subsection{Surface gravity \logg} \label{sec:logg}

The surface gravities, \logg , were evaluated using \textit{Gaia} eDR3 photometry and the standard relation:
\begin{equation}
\log g_{\star}=\log g_{\odot}+\log{\frac{\text{M}_{\star}}{\text{M}_{\odot}}}+  4\log{ \frac{T_{\textsl{eff,}\star}}{T_{\textsl{eff,}\odot}} }+0.4\,(M_{\textsl{bol,}\star}-M_{\textsl{bol,}\odot})
\end{equation}

\noindent Here the distance modulus is $(m-M)_0~=~19.67\pm0.16$ \citep{GaiaHelmi18}, and the mass of the star is assumed to be $\text{M}_\star~=~0.8\pm0.2 $~M$_\odot$. The solar values used are the following: $\log~g_\odot=4.44$, $T_{\textsl{eff,}\odot}=5772$\,K and $M_{\textsl{bol,}\odot}=4.74$. The typical random error is $\Delta\logg=0.14$, arising from error in the $G$ photometry (negligible), \teff, and M$_\star$.

Analogous to \teff, our \logg\ scale is higher than the spectroscopically determined scale. This is a direct consequence of the different 3D NLTE effects of \ion{Fe}{I} and \ion{Fe}{II} lines \citep[e.g.][]{Amarsi16} whose equilibrium is typically enforced by the spectroscopic method. Therefore $\logg-\logg_\textsl{AF}=+0.5$ when compared to \citet{Frebel10}, and $\langle\logg-\logg_\textsl{JS}\rangle=+0.3$ with \citet{Simon15}, while other studies agree on average within $\leq0.06$\,dex \citep{Tafelmeyer10,Starkenburg13,Jablonka15}.

\subsection{The microturbulence velocity}

For the microturbulence velocity, \vt, we use the empirical relation presented in \citet{Kirby09}:

\begin{equation}
\vt=  ((2.13\pm 0.05)-(0.23\pm0.03)\cdot\logg)\, \kms  
\end{equation}
The typical error is $\Delta\vt=0.1\,\kms$, as derived from the errors included in the equation, along with the error on $\log g$ (Sec.~\ref{sec:logg}). This relationship is well tested along a wide range of [Fe/H], and has shown not to produce any significant trends with equivalent width \citep[e.g.][]{Reichert20,Lucchesi2020}, and none were found in a specific test we did for AS0039, the star with the lowest metallicity.

\section{Chemical abundance analysis}

Our stellar sample includes 13 stars in the Sculptor dSph galaxy at $\rm[Fe/H]\leq-2.8$. However, two out of these, DR20080 and ES03170, do not have HR spectra available, and for those stars we only measure Fe and C (along with Ba) to increase the statistics for the fraction of CEMP-no stars in Sculptor at low [Fe/H]. For the remaining 11 stars with HR spectra, we perform a full abundance analysis. All measured abundances are listed in Table~\ref{tab:abundances}. The solar abundances were adopted from \citet{Asplund21}, and all literature data discussed and shown in this paper have been put on the same scale.

\subsection{Stellar atmosphere models and linelists}

The stellar atmosphere models are adopted from MARCS\footnote{\url {marcs.astro.uu.se}} \citep{Gustafsson08} for stars with standard composition, 1D, and assuming local thermodynamic equilibrium (LTE), interpolated to match the stellar parameters for the target stars. The abundance analysis was made with the spectral synthesis code TURBOSPEC\footnote{\url{ascl.net/1205.004}} \citep{AlvarezPlez98,Plez12}, and all individual measurements take blends of other elements into account. Atomic parameters were adopted from the VALD\footnote{\url{http://vald.astro.uu.se}} database (\citealt{Kupka99}), retrieved in October 2019. Finally, the molecular list for CH is obtained from \citet{Masseron14}.

%The errors were evaluated as is described in \citet{Skuladottir17}. When four or more lines were measured for a given element in a star, the final abundance was defined as the average, with the error of the mean. On the other hand, when only three or fewer lines were available, the abundance was defined as the weighted average, and the errors were weighted accordingly (for details, see \citealt{Skuladottir15a}). The systematic errors from the stellar parameters were not included, but in almost all cases, the error on the abundance ratio [X/Y] of two metals is dominated by the line measurement error. For a representative error on [X/Y] due to uncertainties in stellar parameters, see \citet{Hill18}.

\subsection{Chemical abundance measurements}

Chemical abundances and/or upper limits were measured for 2~elements from the LR spectra (2 stars): Fe, C, along with an upper limit of Ba for DR20080; and up to 16 elements from the HR spectra (11 stars): Fe, C, Mg, Na, Al, Si, Ca, Sc, Ti, Cr, Mn, Co, Ni, Zn, Sr, and Ba. Typical errors arising from the uncertainties in stellar parameters were estimated based on a representative star for the sample: $\rm[Fe/H]=-3.4$; $T_\textsl{eff}=4650\,K$; $\logg=1.2$; and $v_t=1.9$\,km/s. Individual components of the error were estimated from synthetic spectra, as arising from $\Delta T_\textsl{eff}=86$\,K, $\Delta\logg=0.14$ and $v_t=0.1$\,km/s, and then combined quadratically. Thus we estimated $\rm\Delta[Fe/H]=0.11$, and $\rm\Delta[X/Fe]\approx 0.1-0.10$ for individual elements measured from atomic lines, $\rm\Delta[X/Fe]$, while $\rm\Delta[C/Fe]=0.15$. The error resulting from stellar parameters was added quadratically to the random observational errors. All measured abundances, upper limits, and uncertainties are listed in Table~\ref{tab:abundances}.

\begin{figure}
\centering
\includegraphics[width=0.95\linewidth]{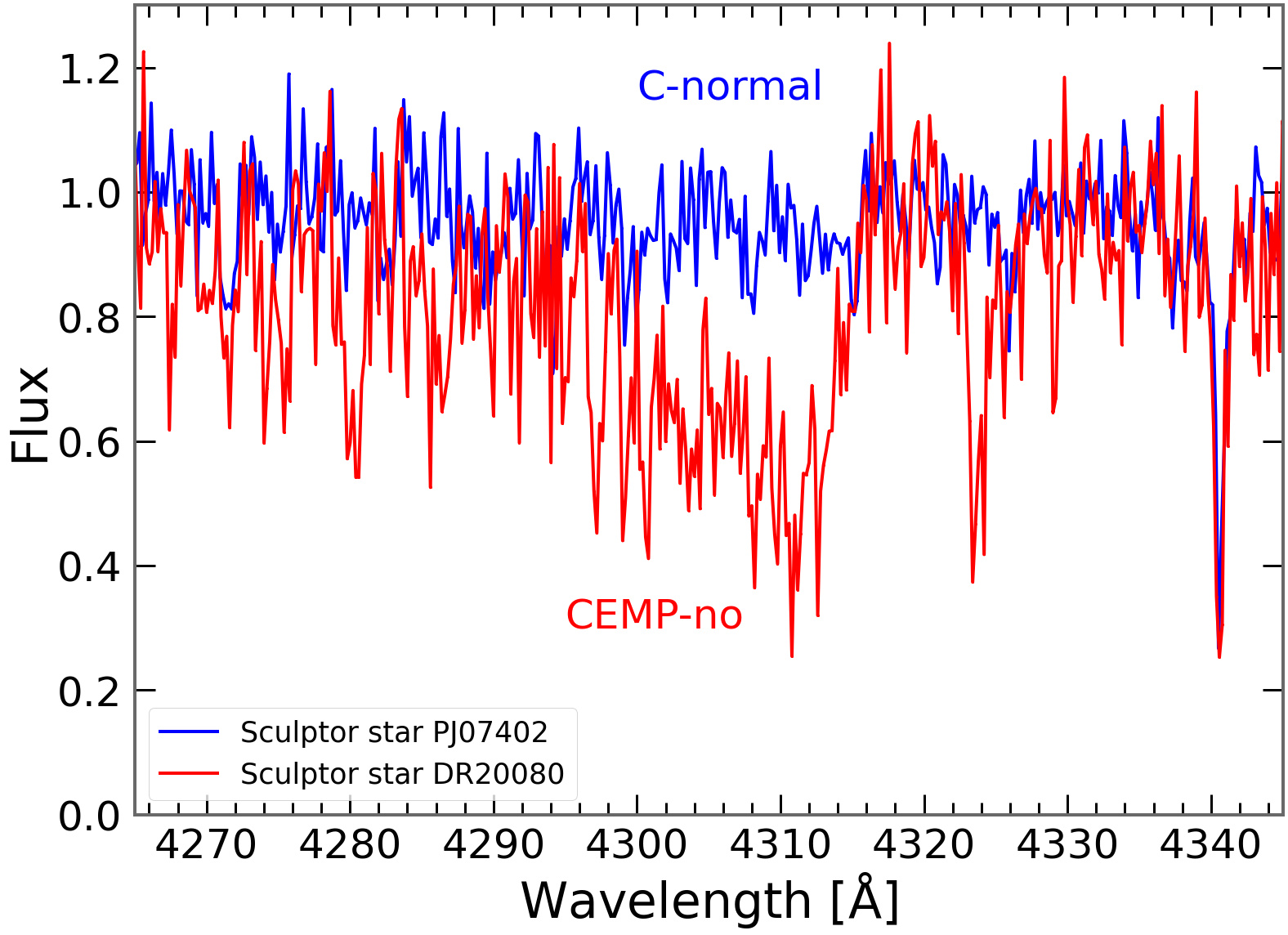}
\caption{X-Shooter spectra for the CEMP-no star in our sample, DR20080 (red), at $\rm[Fe/H]=-3$. For a fair comparison the X-Shooter spectra of PJ07402 at the same $\rm[Fe/H]=-3$, is also shown in blue \citep{Starkenburg13}. However, for the full chemical abundance analysis of PJ07402, here we use a HR UVES spectra from \citet{Jablonka15}.}
\label{fig:dr20080}
\end{figure}

The C abundance was measured from the CH G-band at $\sim$430\,nm in 12 out of our 13 stars (MT00749 lacked the spectral coverage). The strength of these molecular lines are affected by the assumed O abundance, which influences how much of C is locked into CO. The available spectra did not allow us to reliably measure the O abundance, hence we adopt $\rm [O/Fe]=+0.6$, corresponding to the average value at $\rm -2.5\leq[Fe/H]\leq-2$ in Sculptor \citep{Hill19}, and this is also in agreement with the Milky Way halo \citep{Amarsi19}. A combination of low C, high $\teff>4750$\,K, and low S/N lead to only upper limits for the stars AF20549, JS14296, and JS66402. Fig.~\ref{fig:dr20080} shows the spectra of the only CEMP star in the sample, DR20080, around the G-band, comparing it to a C-normal star.

The light, odd element Na was measured with the \ion{Na}{I} D resonance lines at 589.0, and 589.6\,nm, while the Al abundance came from one blue neutral line at 396.2 nm. Both elements were measured in 9 stars, but spectral coverage was missing for two stars in each case.

\begin{figure}
\centering
\includegraphics[width=0.45\linewidth]{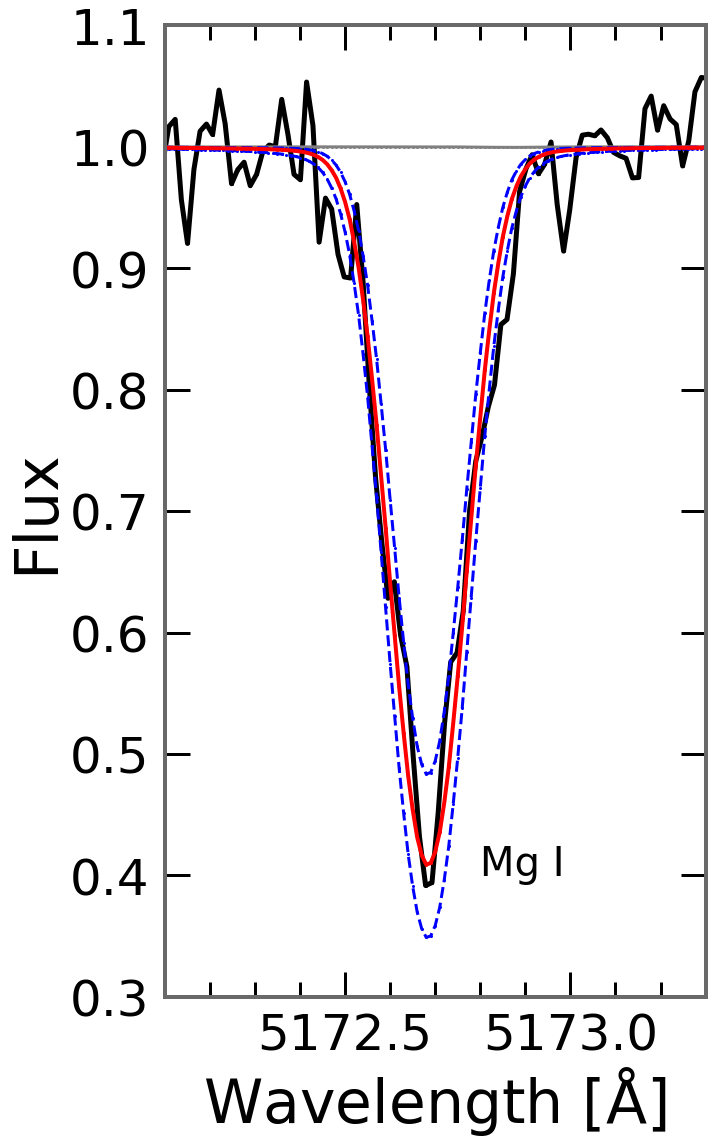}
\includegraphics[width=0.45\linewidth]{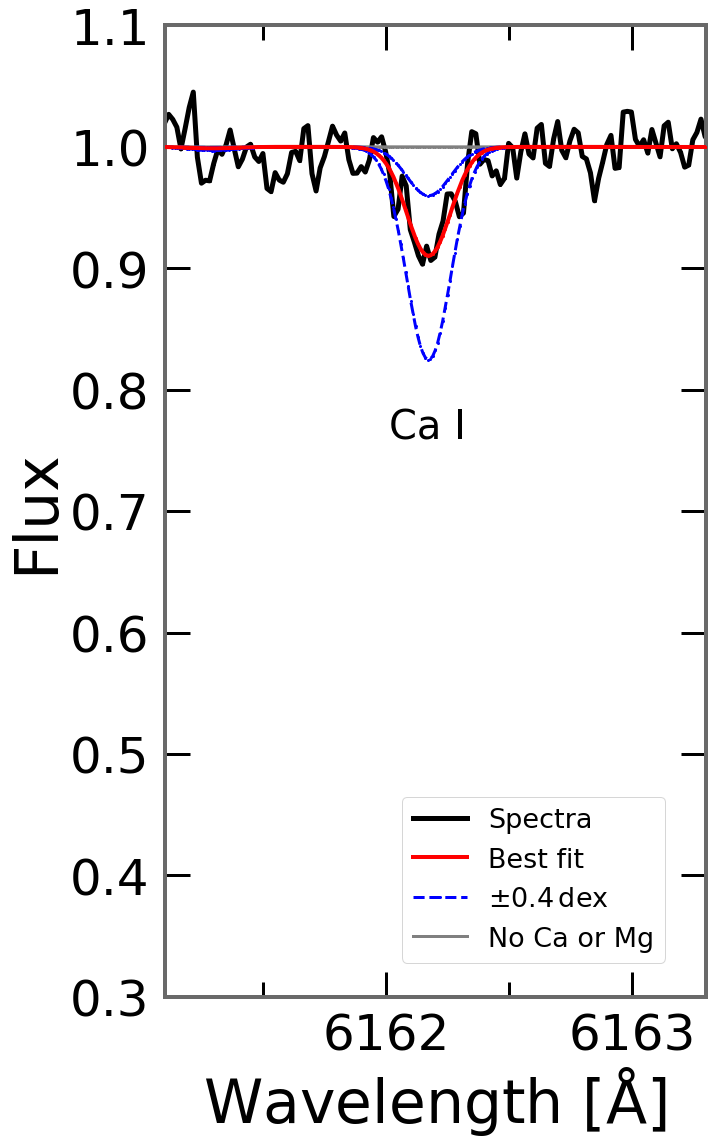}
\caption{Spectrum of AS0039 (black), showing lines of \ion{Mg}{I} (left), and \ion{Ca}{I} (right). Red shows the best fit, blue is $\pm0.4$ from that, and gray shows synthetic spectra without the elements in question.}
\label{fig:mgca}
\end{figure}

Elemental abundances for three $\alpha$-elements were measured, Mg, Si and Ca, as well as the often grouped-together element Ti. We used two \ion{Mg}{I} lines in the blue at 382.8 and 383.8\,nm, and the \ion{Mg}{I} triplet at 516.7, 517.2, and 518.3\,nm. The Si abundance was determined from one neutral line at 390.6\,nm, while five \ion{Ca}{I} lines were used: 422.7, 430.3, 445.5, 612.2, and 616.2\,nm. Finally, Ti was measured by 28 \ion{Ti}{II} lines, ranging from 370-530\,nm. In Fig.~\ref{fig:mgca} the spectral fits are shown for two lines (\ion{Mg}{I} and \ion{Ca}{I}) for the new UVES spectrum of AS0039.

\begin{figure}
\centering
\includegraphics[width=0.90\linewidth]{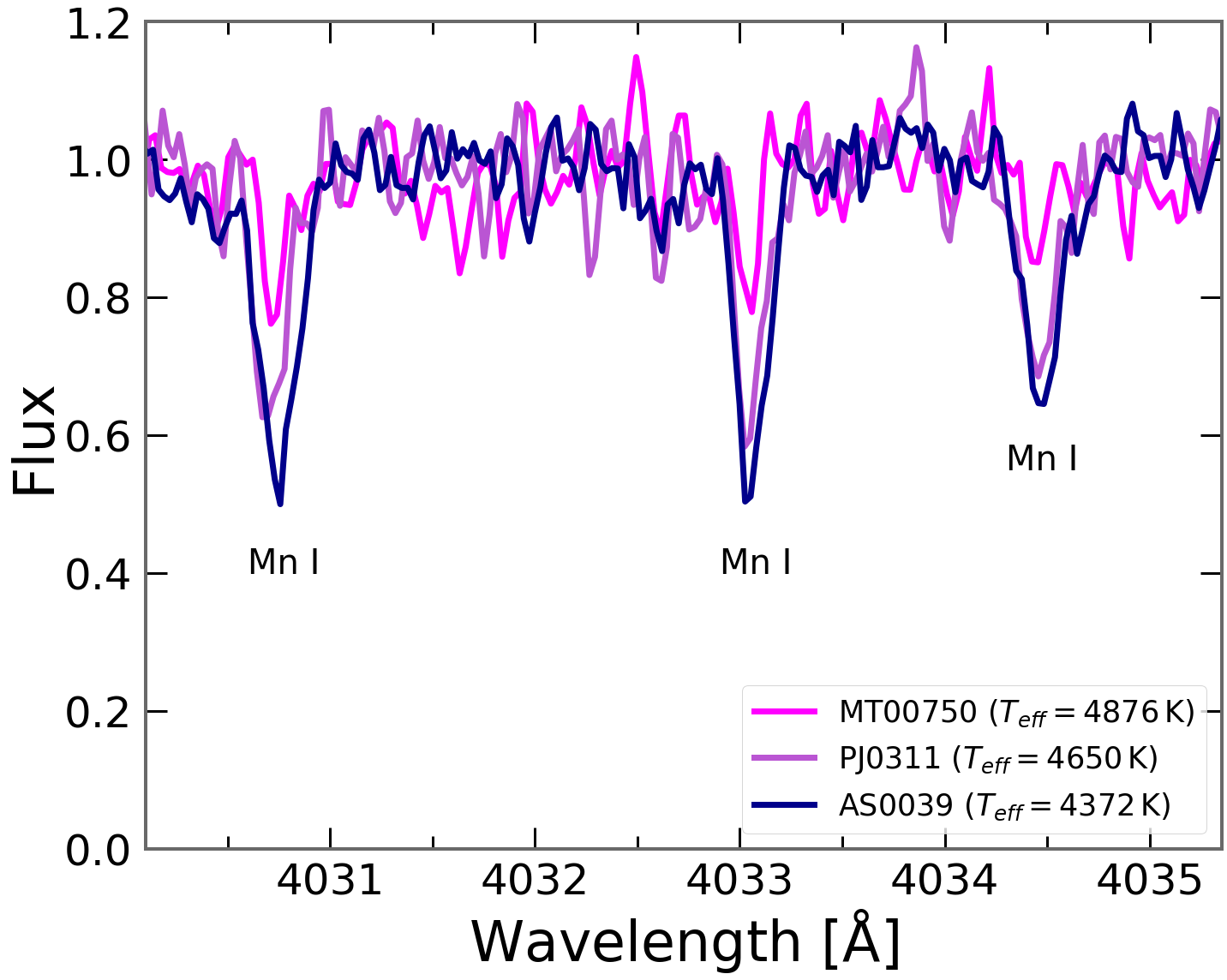}
\caption{Spectra of the strong resonance Mn I triplet for the three most metal-poor stars in our sample, $\rm [Fe/H]\approx-3.85$.}
\label{fig:mntriplet}
\end{figure}

For the odd iron peak elements, we measured Sc using four \ion{Sc}{II} lines at 424.7, 431.4, 432.1, and 432.5\,nm. The Mn was measured from the \ion{Mn}{I} triplet at 403\,nm, shown in Fig.~\ref{fig:mntriplet} for the three most metal-poor stars in our sample. Using eight weak \ion{Co}{I} lines from 384 to 413\,nm we were able to measure Co in 7 of our stars. Due to lack of blue spectral coverage, we were unable to measure Sc, Mn or Co in MT00749 and UHAL004. In addition, the low S/N in JS14296 and JS66402 prevented us from making reliable Co measurements, and the Co in AF20549 is only based on one line (413\,nm). 

We measured four even iron-peak elements: Cr, Fe, Ni, and Zn. In total six \ion{Cr}{I} lines were used for the measurement of Cr, at 425.4, 427.5, 429.0, and the triplet at 520\,nm. The Fe abundance was measured from up to 61 \ion{Fe}{I} lines in the wavelength range $400-570\,$nm, as this was the range covered by most spectra. For the intermediate-resolution X-Shooter spectra (DR20080, and ES03170) the same linelist was used. However due to the lower resolution, some of the weaker lines were not reliably measureable in these stars, though this was somewhat compensated by the above average $\rm[Fe/H]=-3$, compared to the rest of the sample. Thus for DR20080, [Fe/H] was measured using 29 \ion{Fe}{I} lines, while ES03170 was based on 35 lines (this star had spectra with higher S/N). Up to six \ion{Ni}{I} lines were used, at 380.7, 383.2, 385.8, 391.3, 397.4 and 547.7\,nm. For the stars JS14296, JS66402, MT00749, and UHAL004 the Ni abundance is only based on the reddest line at 547.7\,nm. For AF20549 even this line was too weak to measure. Finally, Zn was measured in the 8 stars with the highest S/N, using the line at 481.1\,nm

The neutron-capture elements Sr and Ba were measured, using two \ion{Sr}{II} lines at 407.8 and 421.6\,nm, and up to three \ion{Ba}{II} lines at 455.4, 614.2, and 649.7\,nm. The wavelength range of the bluest line was unfortunately only covered in five stars, AF20549, JS14296, JS66402, MT00749, and MT00750, but all HR spectra included the wavelength covering of the two redder lines. An upper limit of $\rm [Ba/Fe]<-0.5$ was measured for the star DR20080, from the 614.2\,nm line in the X-Shooter spectrum, confirming it to be a CEMP-no star.

\subsection{AS0039 - NLTE abundances}

For detailed comparison with theoretical models (Sec.~\ref{sec:models}), accurate chemical abundances are fundamental. Thus, we provide non-LTE (NLTE) corrections to the abundance measurements of AS0039, the most metal-poor star in our sample. The NLTE chemical abundances for AS0039, along with the size of the NLTE corrections, are listed in Table~\ref{tab:as0039}.

The chemical abundance measurements of \ion{Fe}{I}, \ion{Ca}{I}, and \ion{Ti}{II} were corrected for NLTE effects, on a line-to-line bases, according to \citet{Mashonkina16}, using the publicly available online tool.\footnote{\url{http://spectrum.inasan.ru/nLTE/}} The abundances of Mg, Si, Cr, and Co were corrected with the MPIA-NLTE database\footnote{\url{https://nlte.mpia.de/gui-siuAC_secE.php}} \citep{Bergemann10Cr,Bergemann10Co,Bergemann13,Bergemann17}. The Na correction was derived with the {\tt INSPECT} database\footnote{\url{http://inspect-stars.com/}}, based on the work of \citet{Lind11}, and the Al correction is based on \citet{Nordlander17}. The NLTE correction of Mn was based on the most metal-poor giant models provided by \citet{Bergemann19}. Finally, the Zn NLTE corrections were adopted from \citet{Takeda05}. Corrections of Sc, Sr, and Ba are expected to be small \citep[e.g.][]{Bergemann12,Zhang14} and were neglected here. For the molecular CH lines, 3D corrections are not available but are expected to result in lower abundances \citep{Caffau11,NorrisYong19} compared to the $\rm [C/H]_\text{LTE}$ used in this work. 

When there were several lines of the species available, the NLTE corrections reduced the scatter between lines in all cases except for \ion{Ti}{II}. However, the NLTE corrections for \ion{Ti}{II} are on the order of $\sim$+0.03\,dex, and thus negligible compared to the random errors of singular lines.

\input tables/AS0039.tex

\subsection{Comparison to the literature}

By comparing our results to the previous literature studies of the 9 reanalysed star in our sample (Table~\ref{tab:stellsample}), it becomes clear that our metallicity scale is higher, $\rm \langle[Fe/H]-[Fe/H]_\textsl{lit}\rangle=+0.17$. This is a direct consequence of adopting the photometric \teff\ scale, which is higher than the spectroscopic scale (see Sect.~\ref{sec:stellpar}, and general discussion in \citealt{MucciarelliBonifacio20}). The difference in abundance ratios is smaller, with e.g. $\rm \langle [Mg/Fe]-[Mg/Fe]_\textsl{lit}\rangle=+0.08$. When compared to \citet{Chiti18}, there are five stars in common (AF20549, JS14296, MT00750, PJ03059 and ES03170, respectively 10\_8\_1072, 11\_1\_4296, 10\_7\_923, 10\_8\_320 and 10\_8\_61 in \citealt{Chiti18}), with an average $\rm \langle[Fe/H]-[Fe/H]_\textsl{Chiti}\rangle=+0.01$, and $\sigma=0.17$. 
Out of these five stars, four only have upper limit measurements by us and/or \citet{Chiti18}. For PJ03059, the one star with C measurements by both works, $\rm \langle[C/Fe]-[C/Fe]_\textsl{Chiti}\rangle=-0.36$, which is in agreement within the errors.

The LTE chemical abundances for AS0039 based on the HR UVES spectra are in general agreement with those from the discovery X-Shooter spectrum \citep{Skuladottir21}. In the present work the metallicity is higher, $\rm [Fe/H]_{UVES}-[Fe/H]_{X-Shooter}=+0.21$, with average differences in abundance ratios $\rm \langle[X/Fe]_\textsl{UVES}-[X/Fe]_\textsl{X-Shooter}\rangle=-0.03$, and the differences have an average size of $\rm\langle\,|\,[X/Fe]_\textsl{UVES}-[X/Fe]_\textsl{X-Shooter}\,|\,\rangle=+0.23$. A notable outlier is Ca, with $\rm [Ca/Fe]_{X-Shooter}=+0.65\pm0.11$ and $\rm [Ca/Fe]_\textsl{UVES}=+0.06\pm0.12$. For the weak lines used in both works (422.7 and 616.2\,nm) it is quite likely that unfortunately placed noise affected the X-Shooter measurements. However, these lines were in good agreement with the strong infrared \ion{Ca}{II} triplet (849.9, 854.2 and 866.2\,nm) in the X-Shooter spectra. Unfortunately the UVES spectra does not have the wavelength coverage of the \ion{Ca}{II} triplet, preventing further comparison. With the exception of Ca, the chemical abundances of AS0039 as measured with X-Shooter are consistent with the higher quality UVES analysis.

\begin{figure}

\centering
\includegraphics[width=0.95\linewidth]{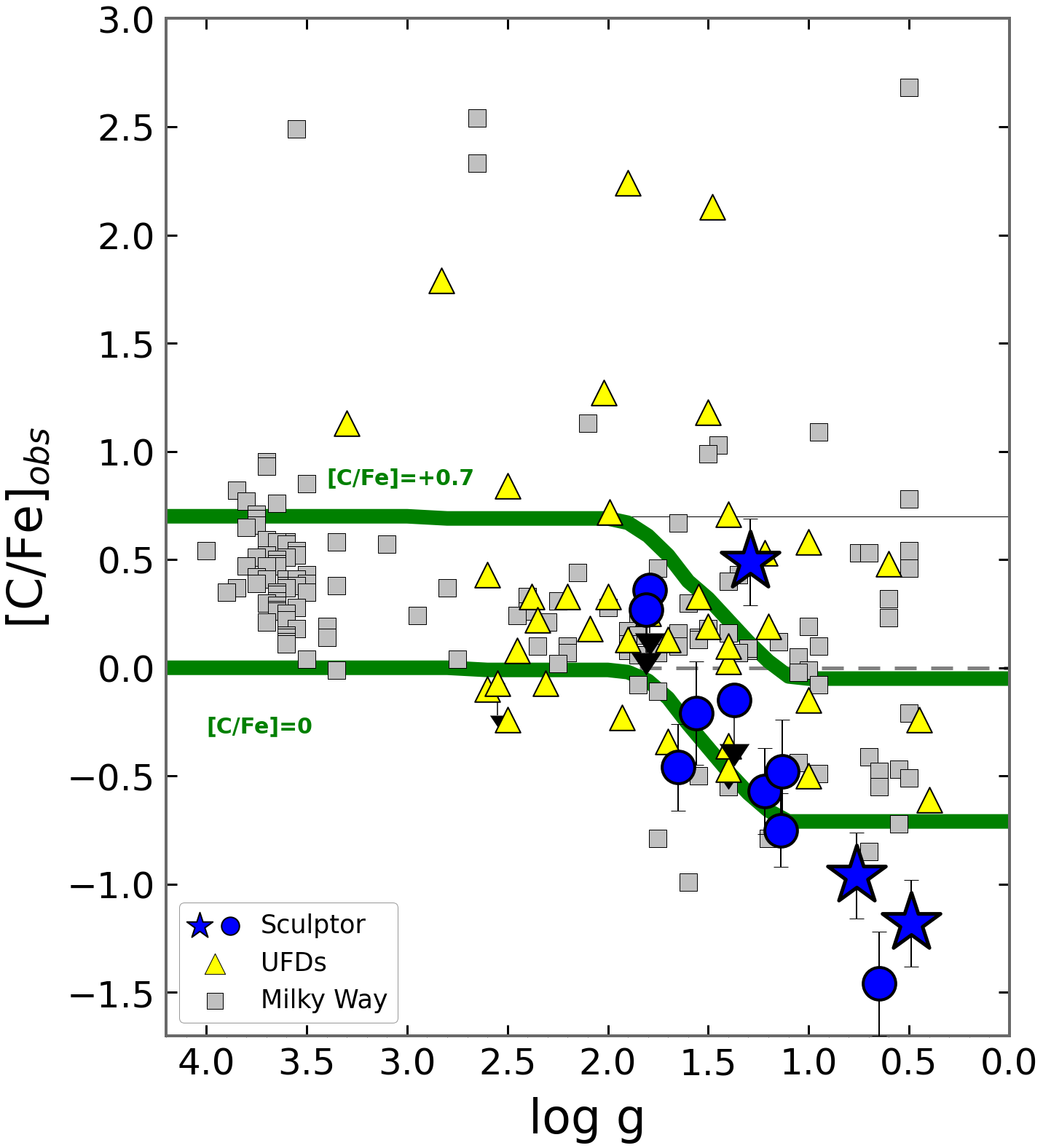}
\par \vspace{0.2cm}
\includegraphics[width=0.95\linewidth]{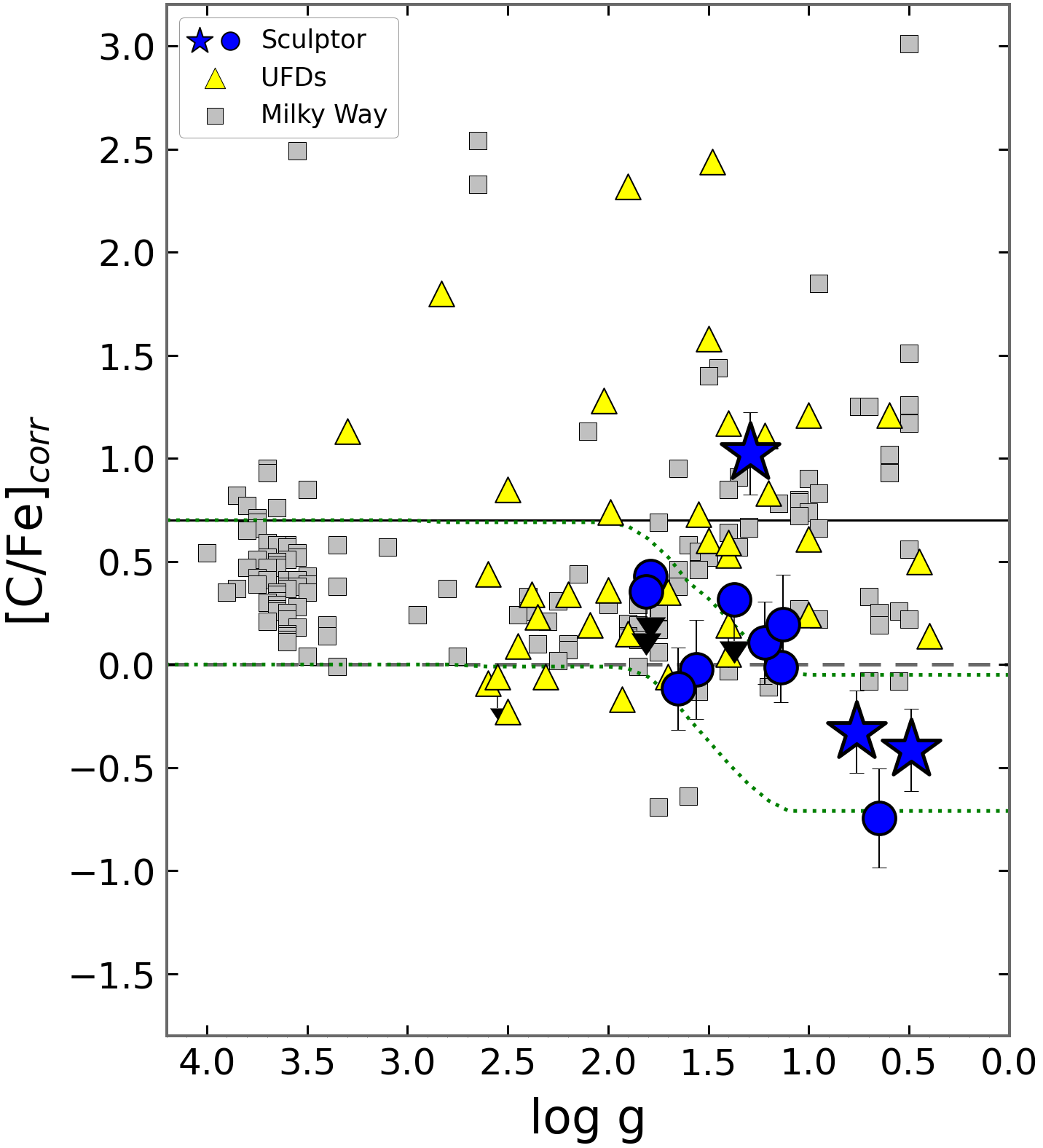}
\caption{Abundance ratios of [C/Fe] with $\log g$ in Sculptor (blue), the Milky Way (gray) and UFDs (yellow), both as measured (top panel), and after applying corrections for internal mixing (bottom panel) according to \citet{Placco14}. Green lines in top panel note the predicted evolutionary track of [C/Fe] in stars born with $\rm [C/Fe]=+0.7$, and 0, repeated with dotted lines in the bottom panel for visual aid. \textit{Milky Way ref.:} \citet{Roederer14}. \textit{UFD ref.:} listed in Sec.~\ref{sec:results}.}
\label{fig:CobsCorr}
\end{figure}

\section{The metal-poor chemical abundances in Sculptor} \label{sec:results}

\begin{figure}
\centering
\includegraphics[width=1\linewidth]{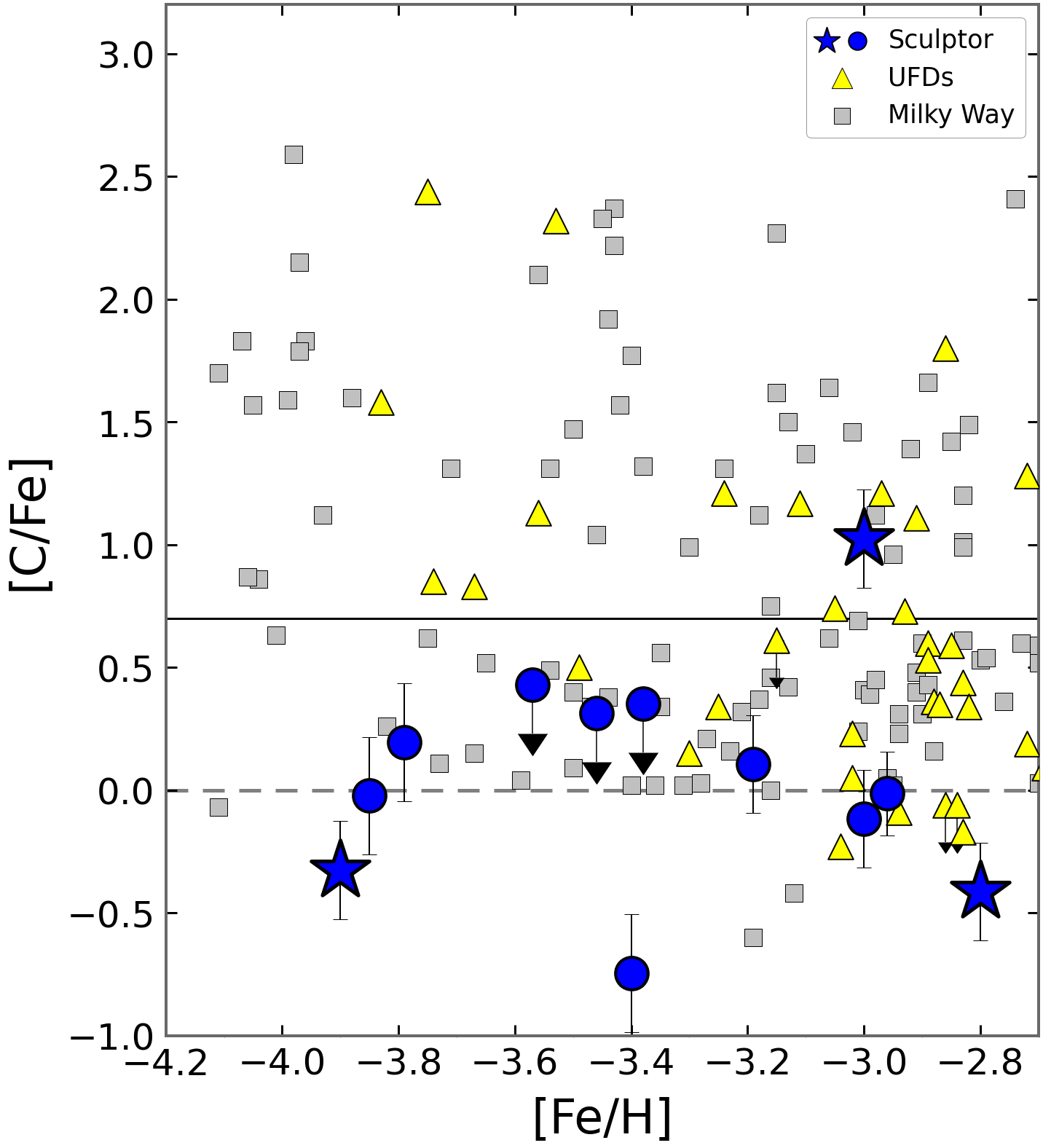}
\caption{Abundance ratios of [C/Fe] with [Fe/H] in Sculptor (blue stars/circles), the Milky Way (gray squares) and UFDs (yellow stars), corrected for internal mixing \citep{Placco14}, excluding CEMP-$s$ stars. Ref. listed in Sec.~\ref{sec:results}. }
\label{fig:C}
\end{figure}

\begin{figure*}[ht]
\centering
\includegraphics[width=0.42\linewidth]{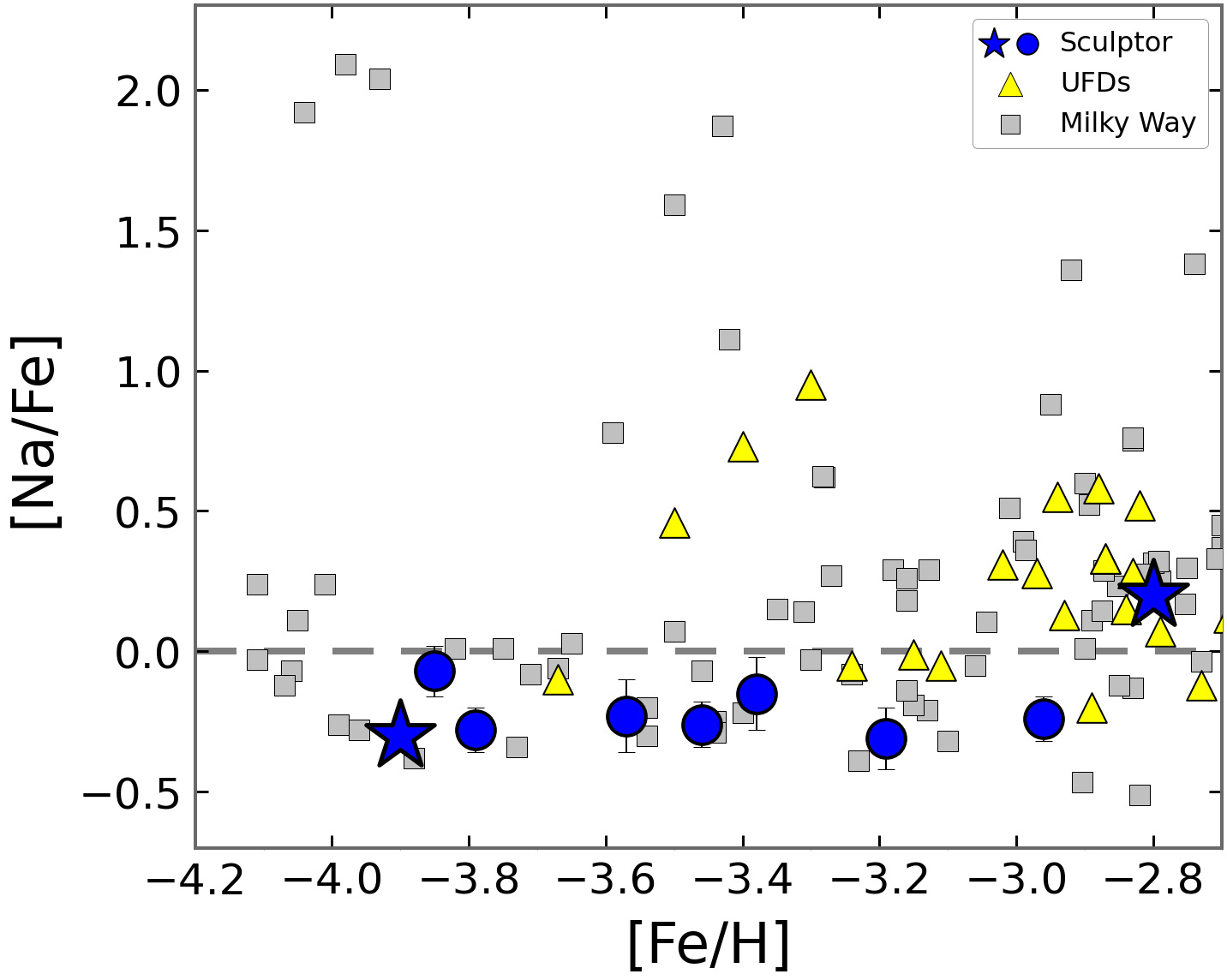}
\hspace{0.5cm}
\includegraphics[width=0.42\linewidth]{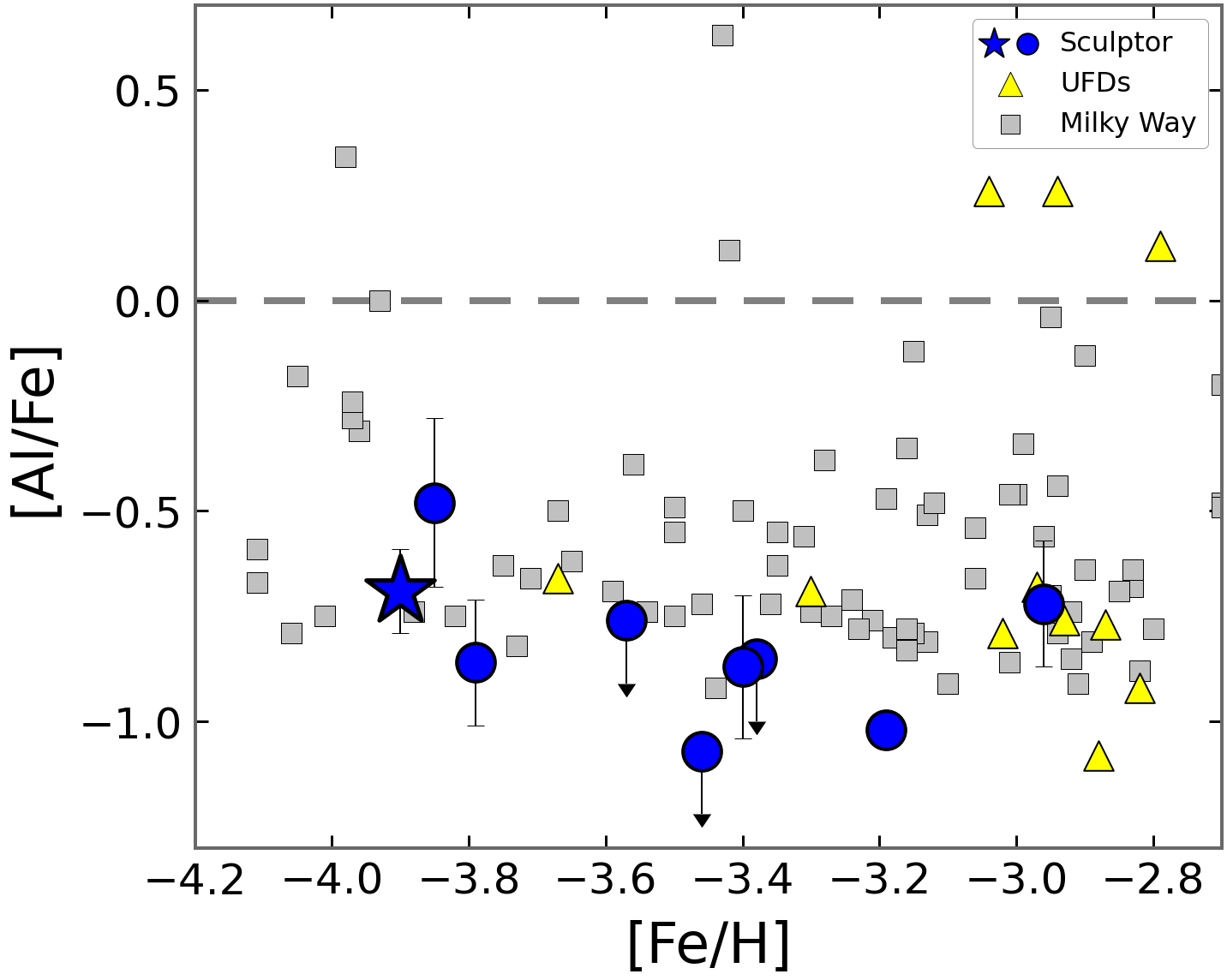}
\caption{The light odd elements, [Na/Fe] and [Al/Fe] with [Fe/H] for Sculptor (blue), the Milky Way (gray), and UFDs (yellow). Star symbols represent Sculptor stars analysed here using new spectra, while blue circles are stars reanalysed on archival spectra.}
\label{fig:NaAl}
\end{figure*}

\begin{figure*}[ht]
\centering
\includegraphics[width=0.42\linewidth]{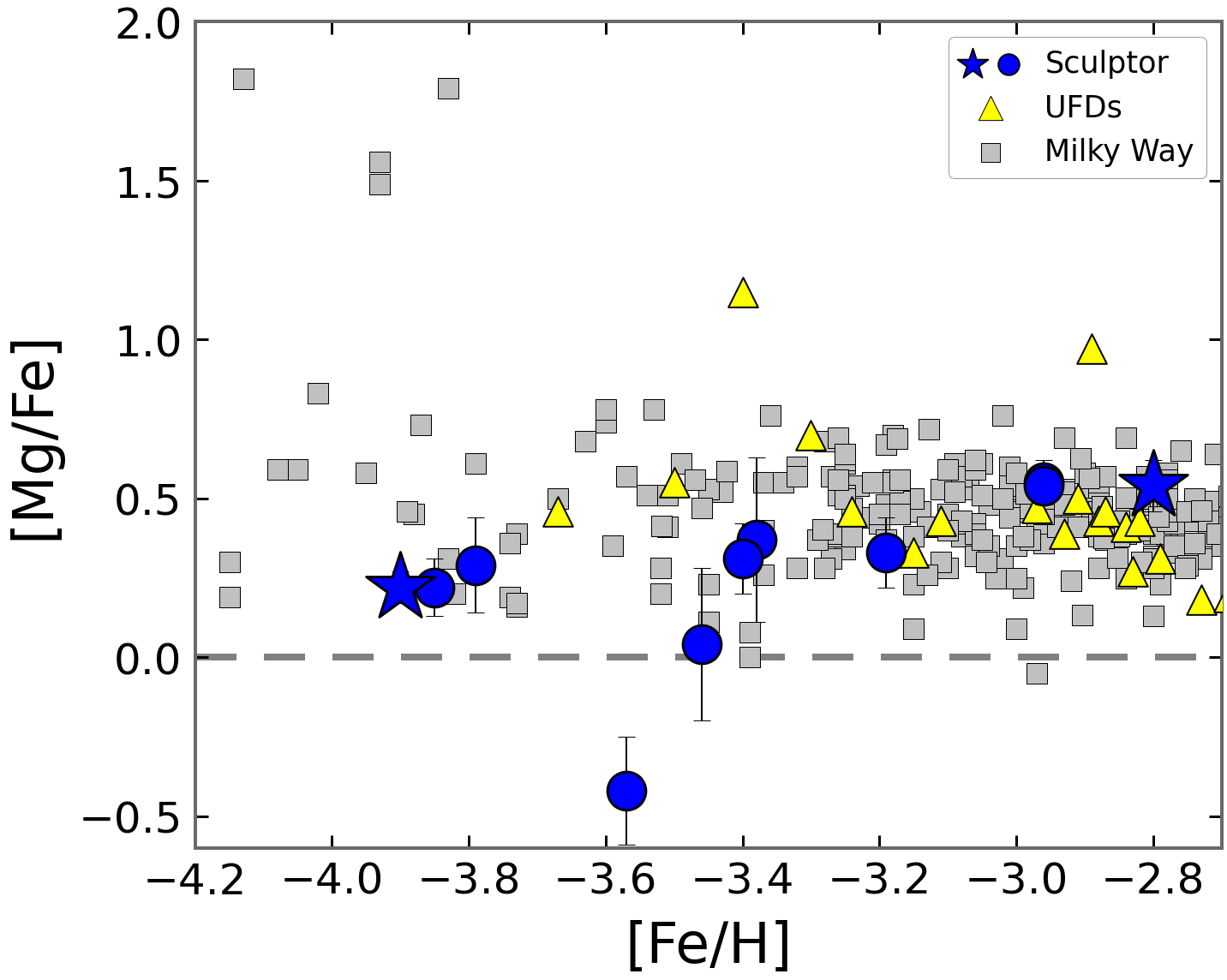}
\hspace{0.5cm}
\includegraphics[width=0.42\linewidth]{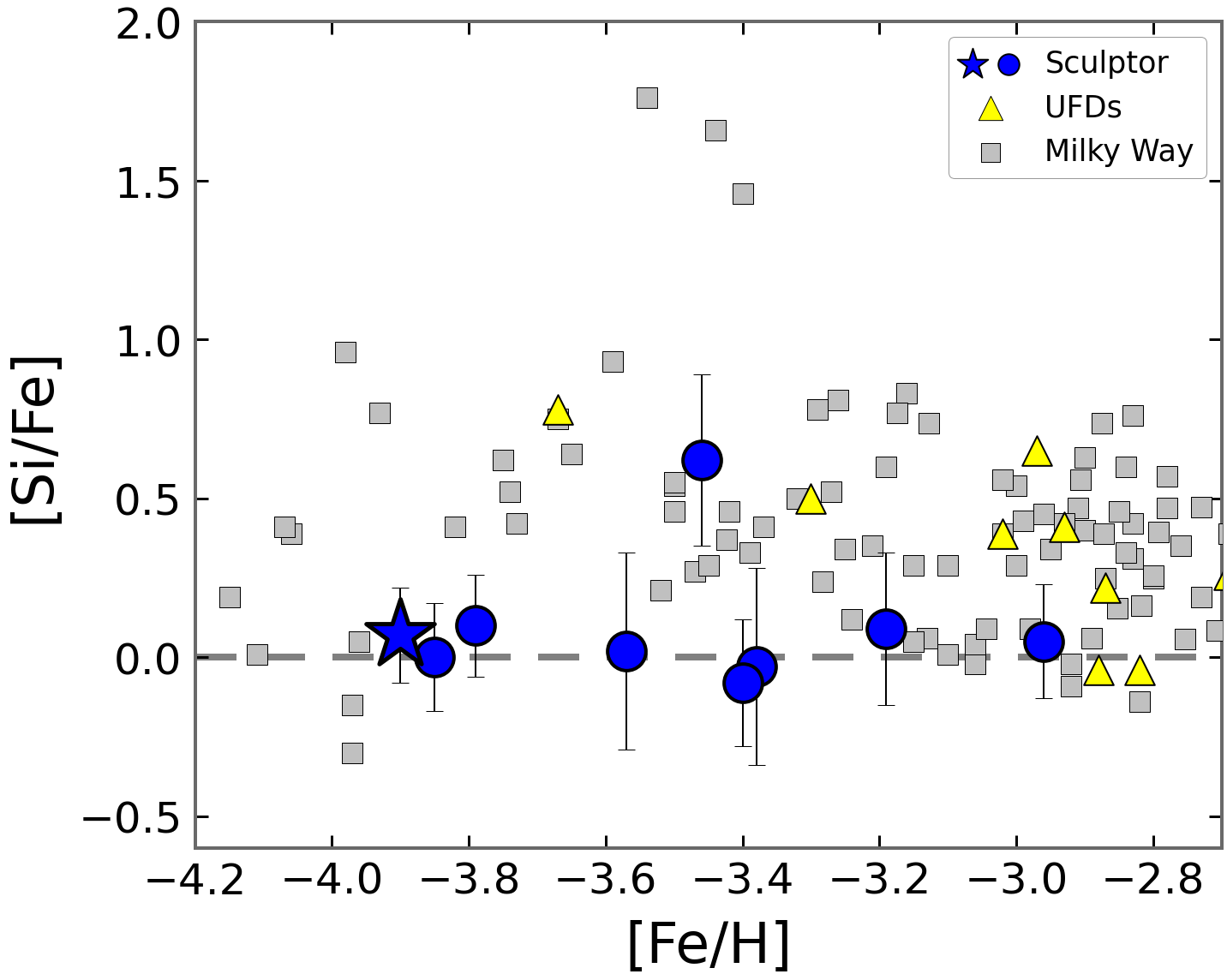}
\par \vspace{0.2cm}
\includegraphics[width=0.42\linewidth]{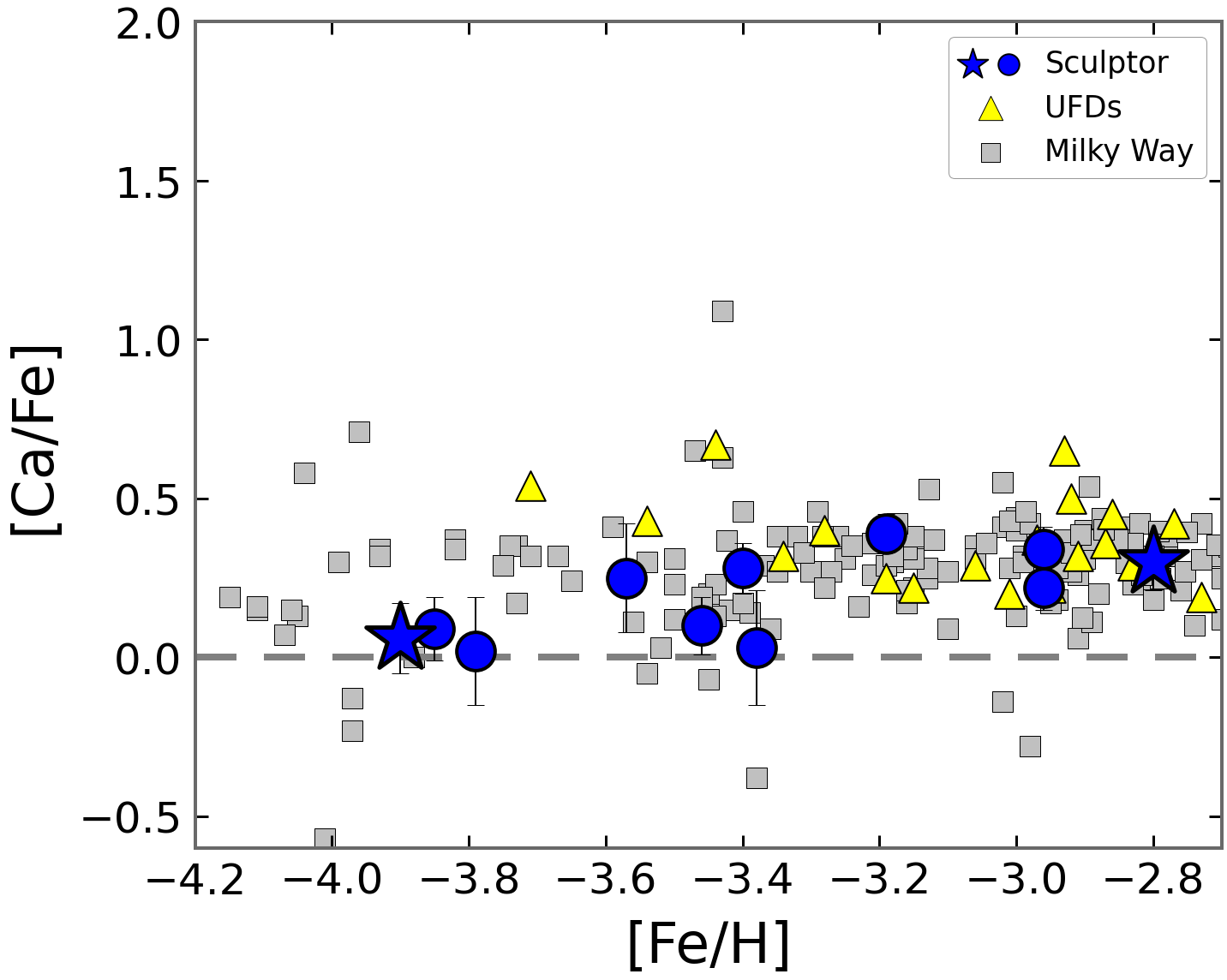}
\hspace{0.5cm}
\includegraphics[width=0.42\linewidth]{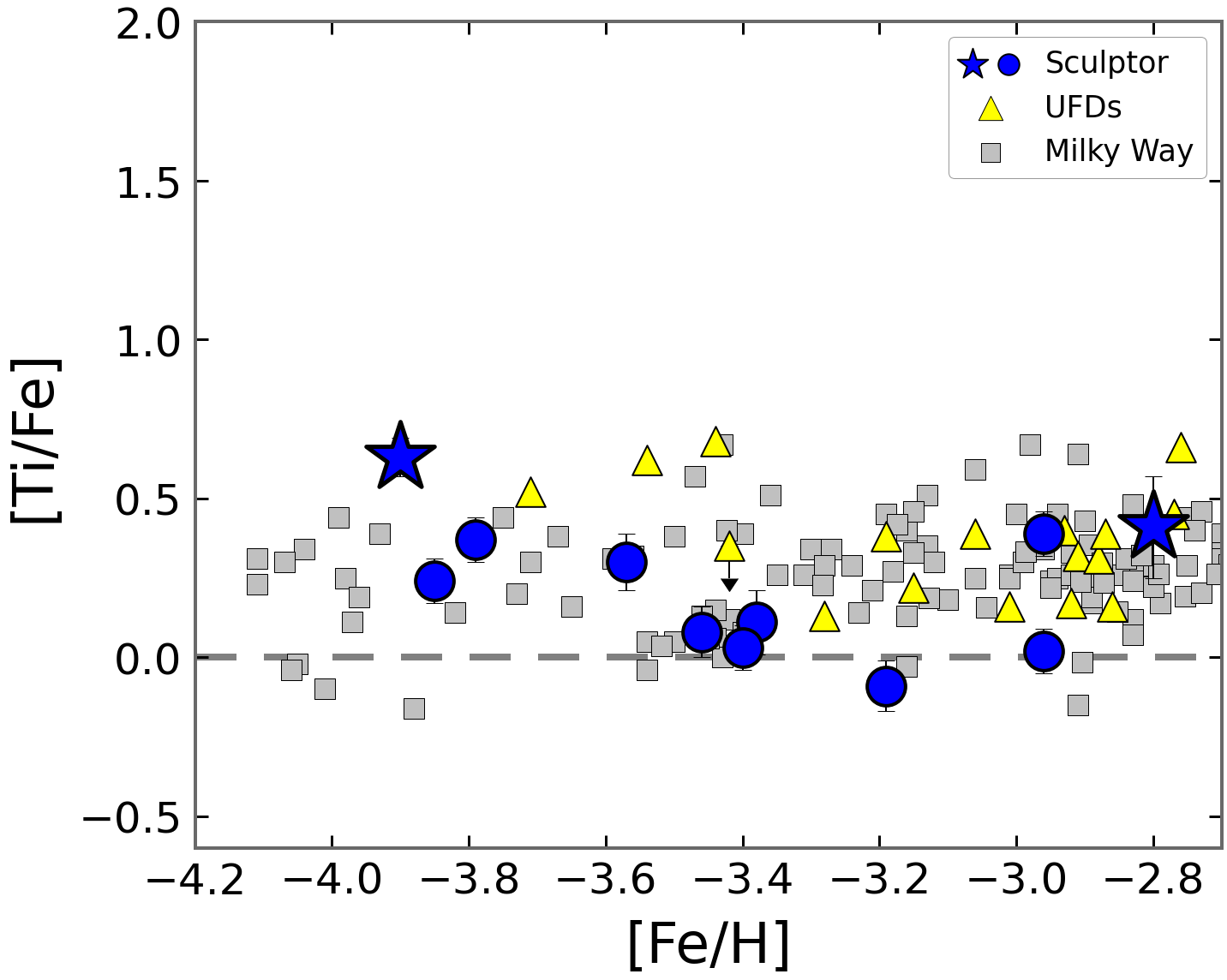}

\caption{The $\alpha$-elements, [Mg/Fe], [Si/Fe], and [Ca/Fe], along with [Ti/Fe], with [Fe/H] for Sculptor (blue), the Milky Way (gray), and UFDs (yellow). Star symbols represent Sculptor stars analysed here using new spectra, while blue circles are stars reanalysed on archival spectra.}
\label{fig:alpha}
\end{figure*}

\begin{figure*}
\centering
\includegraphics[width=0.32\linewidth]{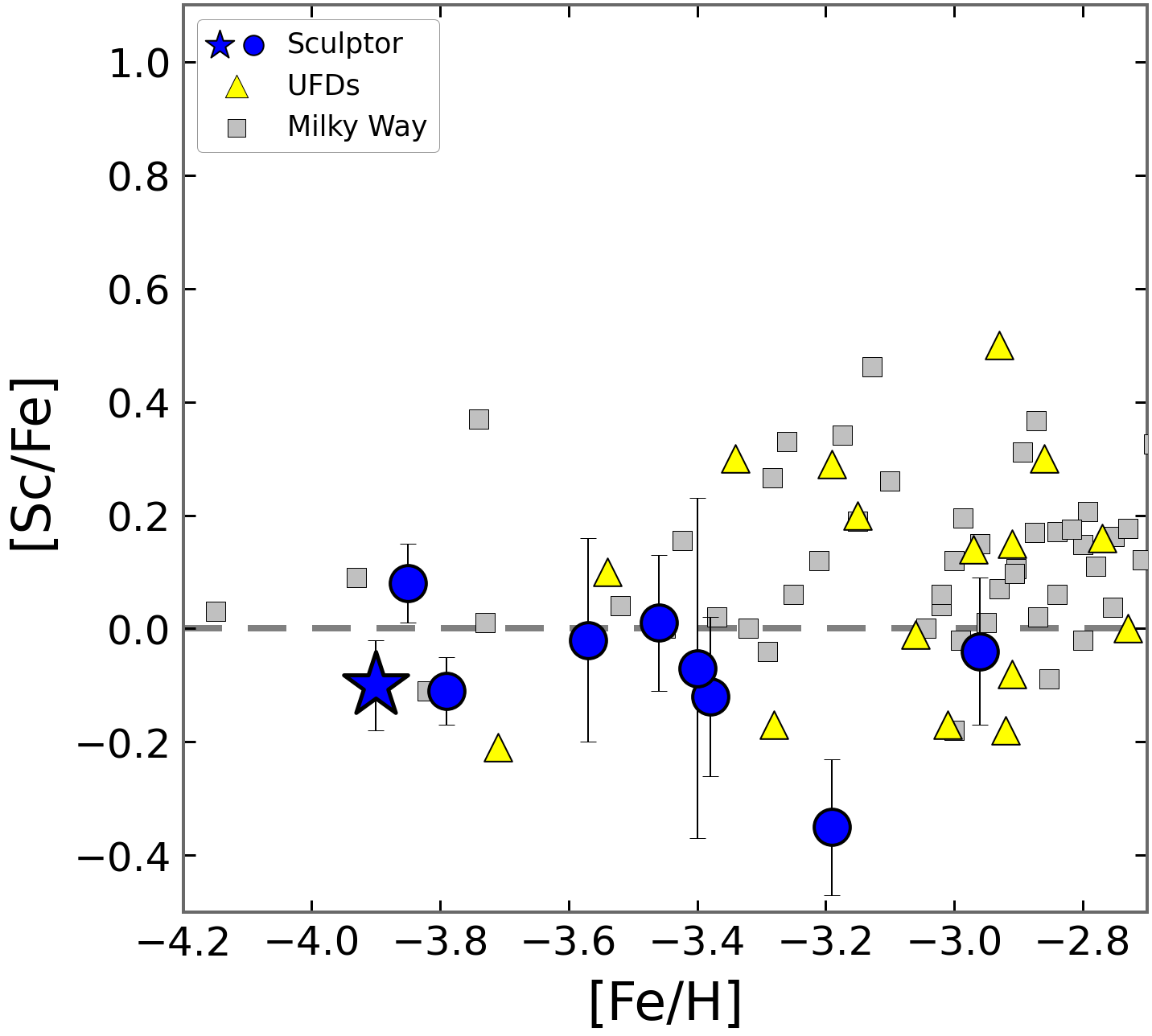}
\hspace{0.2cm}
\includegraphics[width=0.32\linewidth]{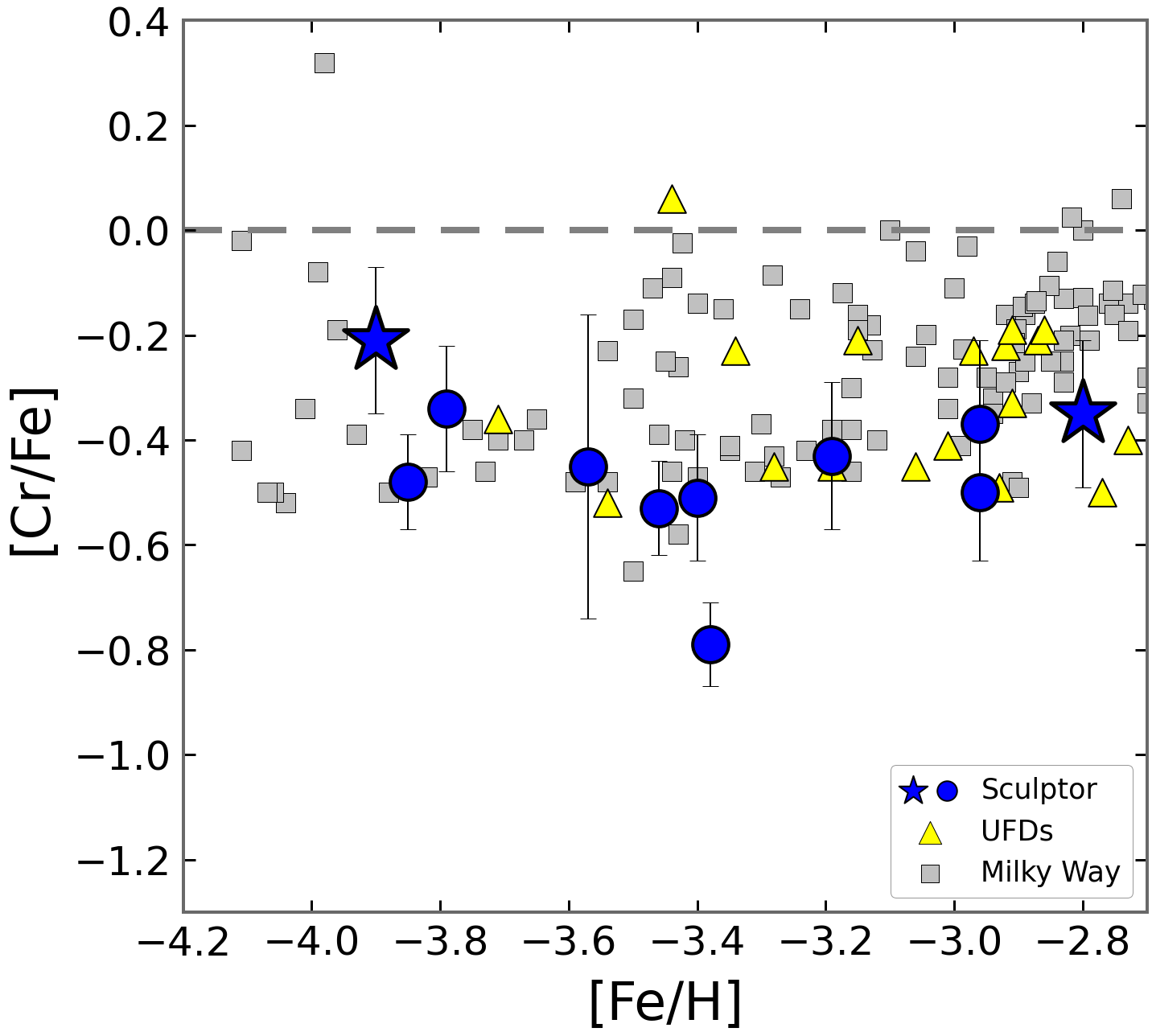}
\hspace{0.2cm}
\includegraphics[width=0.32\linewidth]{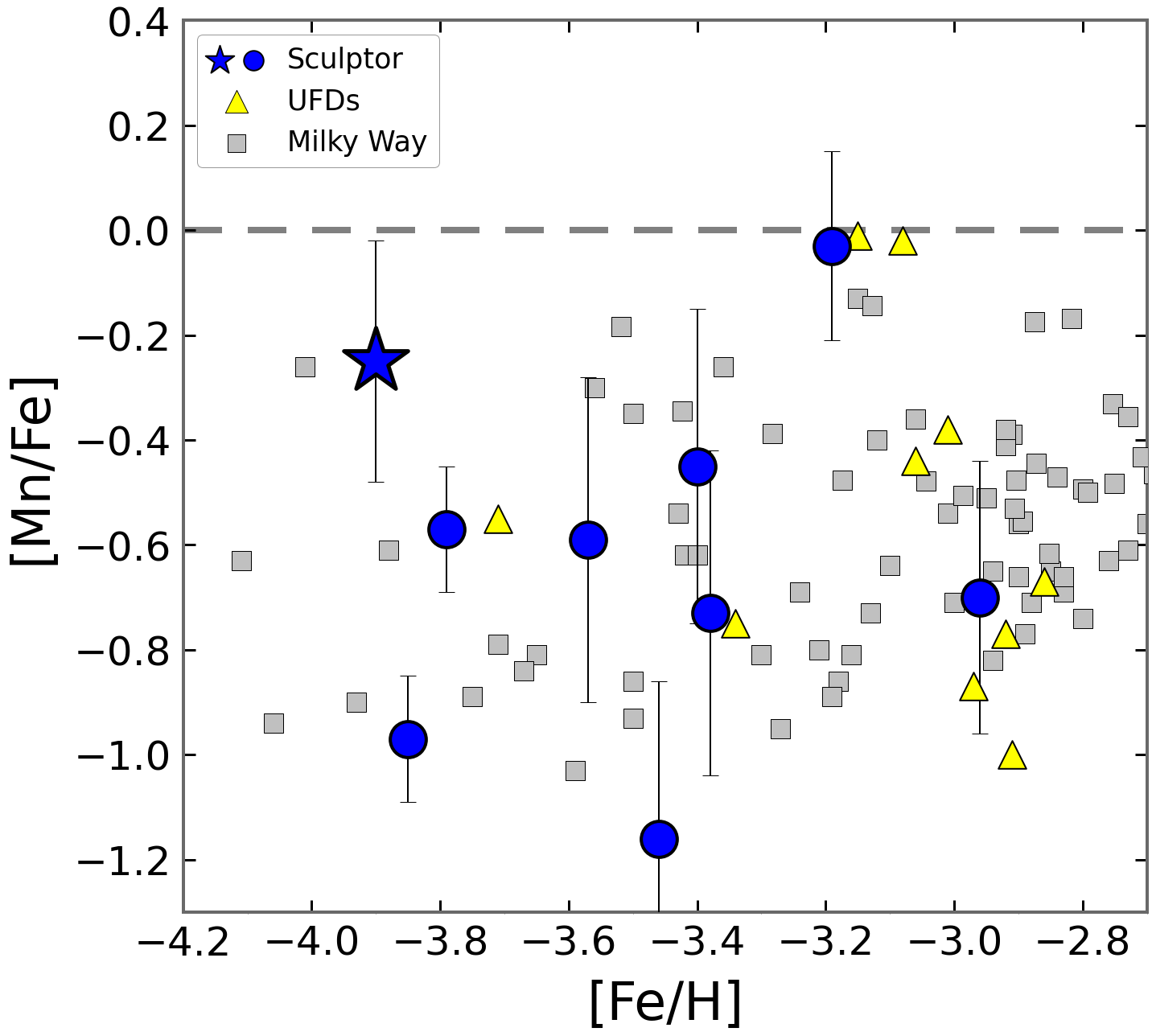}
\par \vspace{0.2cm}
\includegraphics[width=0.32\linewidth]{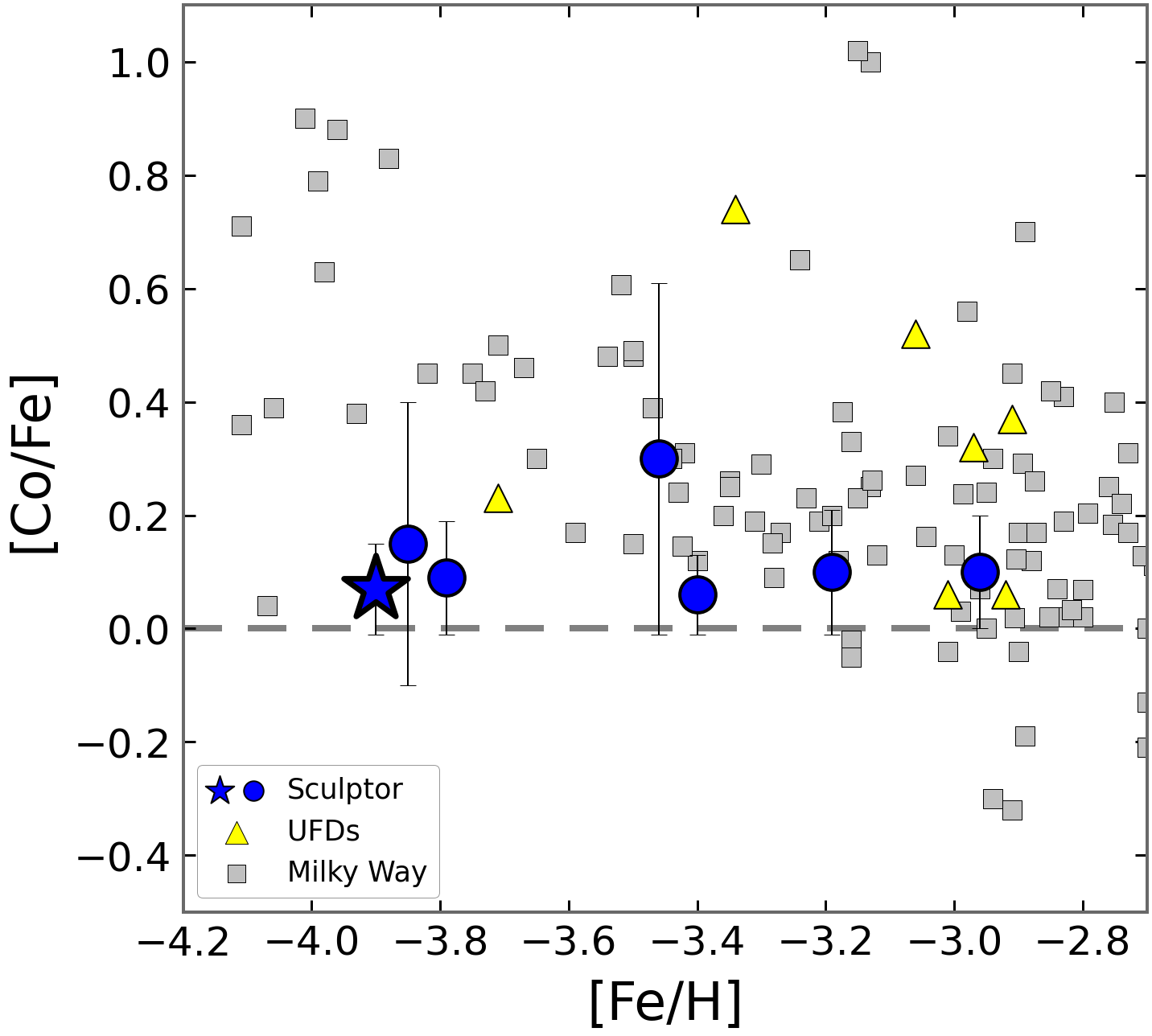}
\hspace{0.2cm}
\includegraphics[width=0.32\linewidth]{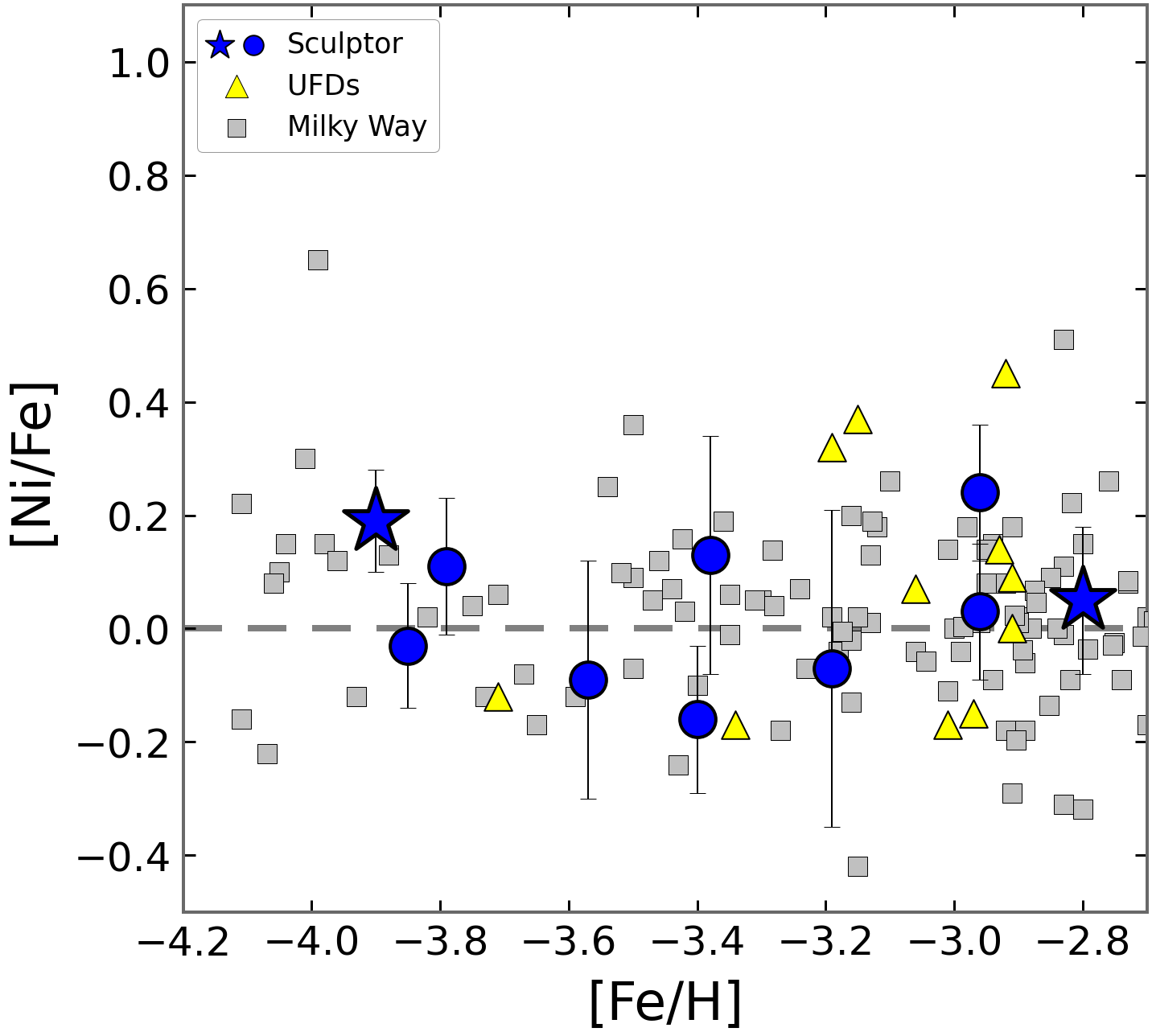}
\hspace{0.2cm}
\includegraphics[width=0.32\linewidth]{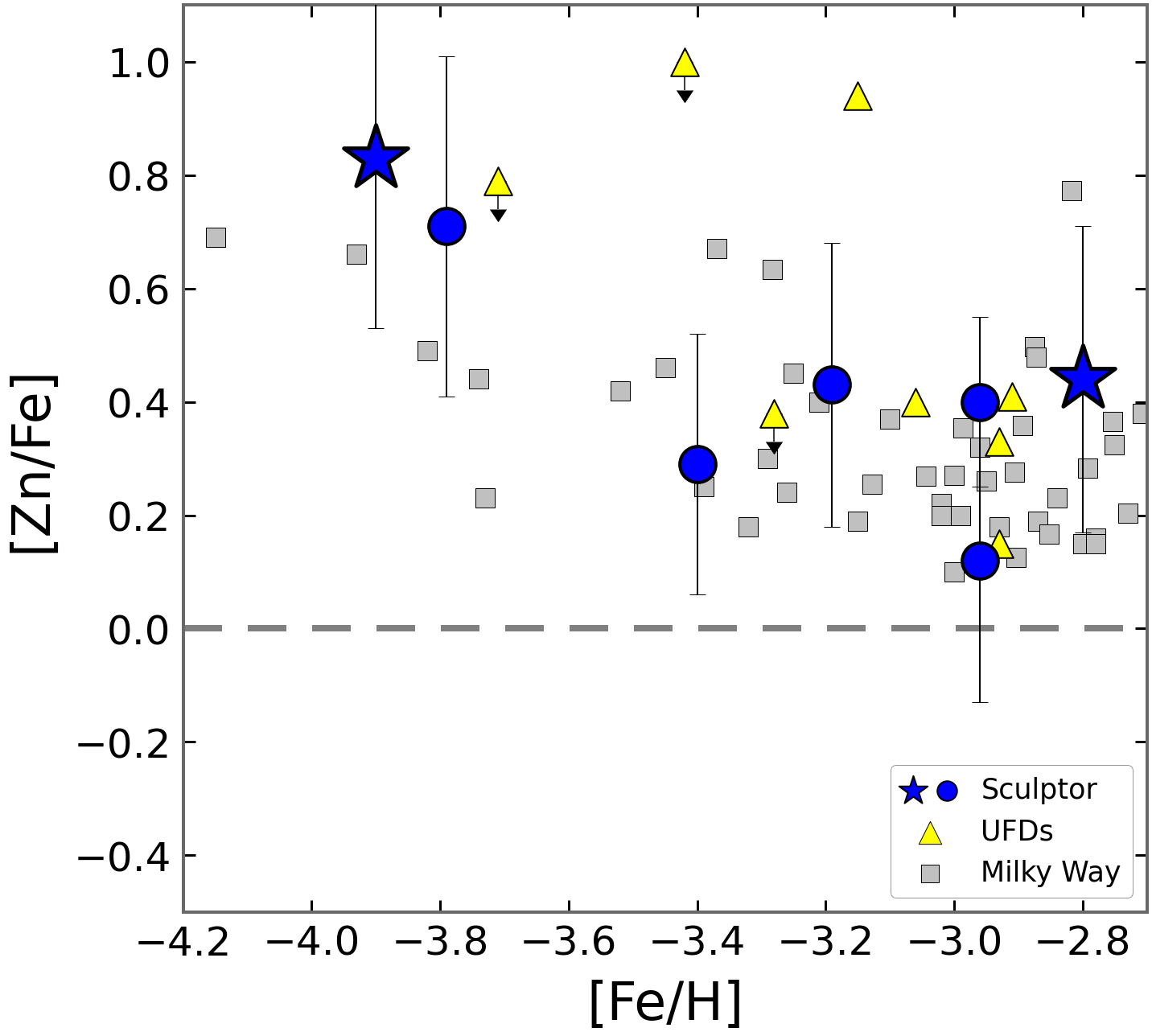}
\caption{The iron-peak elements Sc, Cr, Mn, Co, Ni and Zn relative to Fe, as a function of [Fe/H] for Sculptor (blue), the Milky Way (gray), and UFDs (yellow). Star symbols represent Sculptor stars analysed here using new spectra, while blue circles are stars reanalysed on archival spectra.}
%from metov.py
\label{fig:ironpeak}
\end{figure*}

The chemical abundance measurements (LTE) of our Sculptor stars are listed in Table~\ref{tab:abundances}, and shown in Figs~\ref{fig:CobsCorr}-\ref{fig:pattfig}. In the following subsections we will discuss the results and put them in context with the metal-poor Milky Way halo and the UFDs. In particular, in Sec.~\ref{sec:genabu}, we quantitatively compare the overall abundance patterns in different galaxies.

For the Milky Way we use literature data from: \citet{Cayrel04,Christlieb2004,Norris2007,Caffau2011halo,Yong2013,Hansen2014,Keller2014,Roederer14,Frebel2015,Bonifacio2015,Li2015,Bonifacio2018,Francois2018,Starkenburg2018,Aguado2019,Ezzeddine2019,Lombardo22}. For the UFDs we use data from the JINA database\footnote{\url{https://jinabase.pythonanywhere.com/}}\citep{abohalima2018jinabase}, which includes: \citet{Norris10,Simon10,Frebel10b,Frebel16,Ishigaki14,Roederer16,Francois16,Ji16a,Ji16b}. For C abundances we also include the following UFD references: \citet{Lai11,Ji20}. For duplicate entries of the same stars, the original publication is used. All literature abundances are put on the solar scale of \citet{Asplund21}, see Tables~\ref{tab:abundances} and \ref{tab:as0039}.

\subsection{Carbon} \label{sec:carbon}

The C abundance on the surface of RGB stars is modified by internal dredge-up and mixing \citep[e.g.][]{Gratton00,Spite06}. To correct for this effect, we used the online tool\footnote{\url{http://vplacco.pythonanywhere.com/}} provided by \citet{Placco14}. Fig.~\ref{fig:CobsCorr} shows both the measured abundances, $\rm [C/Fe]_\textsl{obs}$, and those corrected for mixing, $\rm [C/Fe]_\textsl{corr}$ as a function of $\log g$. The average size of the correction for the Sculptor sample is $\rm \langle \,\Delta[C/Fe]\,\rangle=+0.5$. We note that adopting lower $\log g$ would lead to higher corrections. In general our $\log g$ scale is in good agreement with the literature (Sec.~\ref{sec:logg}), with the exception of \citet{Frebel10} and \citet{Simon15}, where $\Delta \log g\gtrsim 0.3$. However for those three stars we were only able to provide upper limits, see Fig.~\ref{fig:CobsCorr}, and our results are in general agreement with \citet{Frebel10} and \citet{Simon15}, which categorised these stars as C-normal. 

In the case of $\rm [C/Fe]_\textsl{obs}$ (Fig.~\ref{fig:CobsCorr}, top) there is a steep negative gradient with evolutionary stage, i.e. decreasing \logg, as is expected from theory. After applying the corrections, the trend is flattened at $\log g<1$. However, we note that the three most evolved stars are still significantly lower in $\rm [C/Fe]_\textsl{corr}$ compared to the other C-normal stars, with $\rm [C/Fe]_\textsl{corr}=-0.5$, while other C-normal stars are consistent with $\rm [C/Fe]_\textsl{corr}\approx0.0$. Given that these three most evolved stars (AS0039, PJ00306, and UHAL004) cover the entire metallicity range of the sample, it is likely that this is an artificial effect, produced either by underestimation of the C-corrections at low $\logg<1$ and/or differences in 3D effects of the CH band. If the corrections are underestimated at low $\log g$, our presented abundances are too low, however, adding 3D corrections to the abundances derived from the G-band, will lower these abundances \citep[e.g.][]{NorrisYong19}. For the lack of a better alternative we accept the abundances as presented, with the aforementioned caveats. In any case, it is unambiguous that these three evolved stars are all C-normal, $\rm[C/Fe]<+0.7$.

Fig.~\ref{fig:C} shows $\rm [C/Fe]_\textsl{corr} $ (from here on labelled [C/Fe]) as a function of [Fe/H]. In the Sculptor sample of 12 stars at $\rm-4<[Fe/H]\leq-2.8$, only one star, DR20080, is carbon-enhanced, $\rm[C/Fe]>+0.7$ (see also Fig.~\ref{fig:dr20080}). This star has $\rm [Ba/Fe]<-0.5$ (Table~\ref{tab:abundances}), and is therefore a bonafide CEMP-no star. This gives a fraction $f_\textsl{CEMP}^\textsl{Scl}=9^{+11}_{-8}\%$ at $\rm[Fe/H]\leq-3.0$, while the fractions measured in the literature in the Milky Way and UFDs are $f_\textsl{CEMP}\approx40\%$ \citep{Placco14,Ji20}. Furthermore we emphasise that at $\rm[Fe/H]<-3.5$ there is no known CEMP-no star in Sculptor, while the fraction in the Milky Way halo and UFDs is $f_\textsl{CEMP}>60\%$. Therefore it is evident that although CEMP-no stars do exist in Sculptor (\citealt{Skuladottir15a}; and here) their fraction is significantly lower in Sculptor compared to the Milky Way halo and the UFDs. This dearth of CEMP-no stars in Sculptor has been noted previously in the literature \citep[e.g.][]{Starkenburg13,Skuladottir15a,Skuladottir21}, and is further solidified with this work. This clearly points to a different chemical enrichment in Sculptor at the earliest epochs, compared to both the Milky Way halo and the UFDs. This is further discussed in Sec.\ref{sec:models}.

\subsection{Light odd elements, Na and Al}

The abundances of the light odd elements Na and Al are shown in Fig.~\ref{fig:NaAl}. Sculptor follows the lower envelope of [Na/Fe] values observed in the Milky Way and UFDs, with all but the most metal-rich star, UHAL004, having subsolar $\rm[Na/Fe]<0$. A similar picture arises with [Al/Fe], where the Sculptor abundances are typically lower than the median of the Milky Way. The UFDs show a bimodal behavior in [Al/Fe], either being similar to Sculptor, or very high with $\rm [Al/Fe]>0$. We note that all the supersolar abundances in UFDs in Fig.~\ref{fig:NaAl} are from the same reference \citep{Francois16}, but the authors find similar results for several UFDs (Boötes~II, Hercules, Leo~IV).

From Fig.~\ref{fig:NaAl} it is clear that Sculptor is significantly lower in the abundances of light odd elements, Na and Al, relative to Fe, compared to other depicted galaxies. The low Na is likely closely related to the dearth of CEMP-no stars in Sculptor, since such stars are known to often also have high Na and Al \citep[e.g.][]{Norris13}. But even when only considering C-normal RGB stars, the Sculptor values are typically lower than the median of Milky Way data, suggesting an intrinsic lack of these elements in the earliest chemical enrichment of Sculptor.

\begin{figure*}[ht]
\centering
\includegraphics[width=0.32\linewidth]{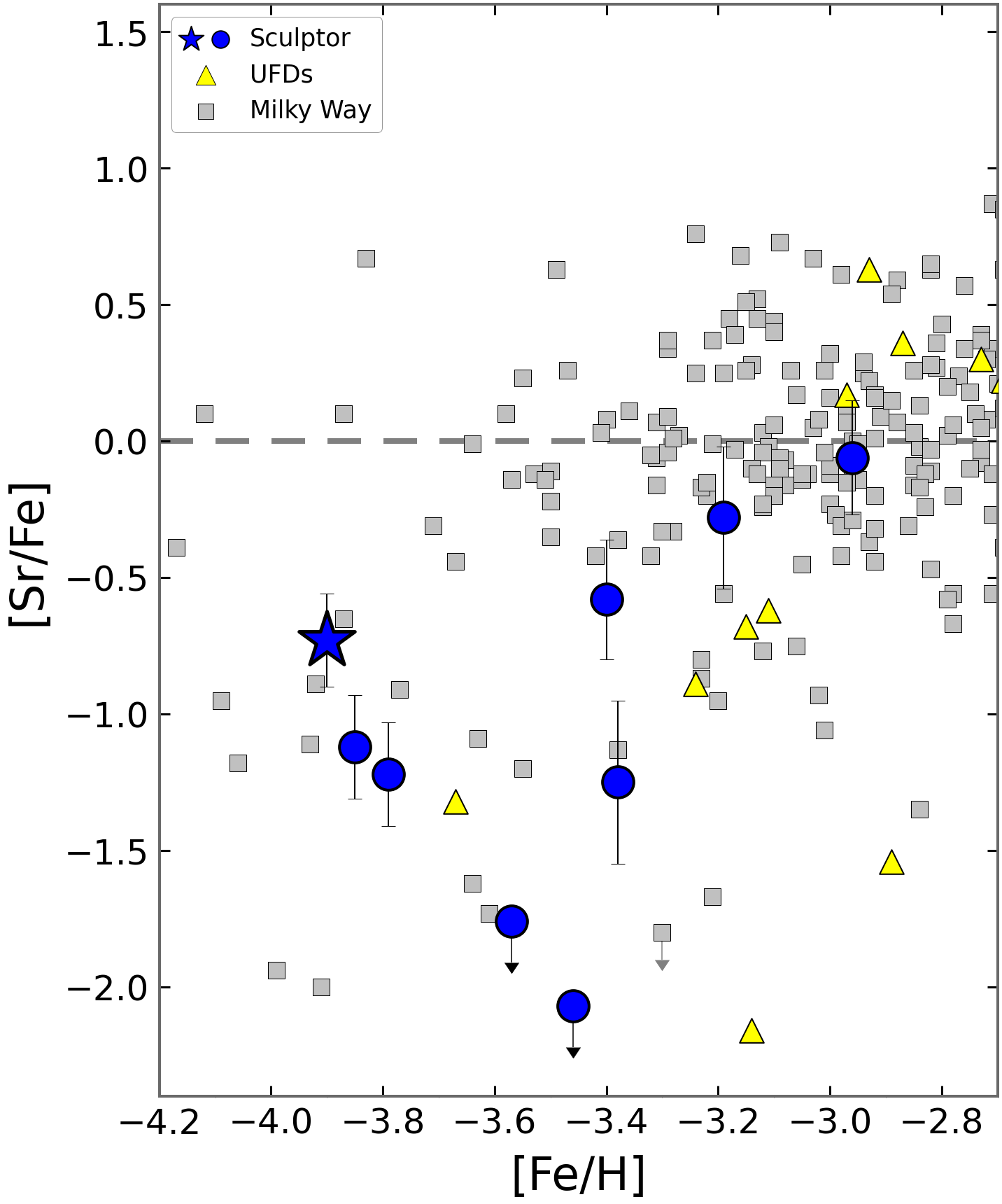}
\hspace{0.2cm}
\includegraphics[width=0.32\linewidth]{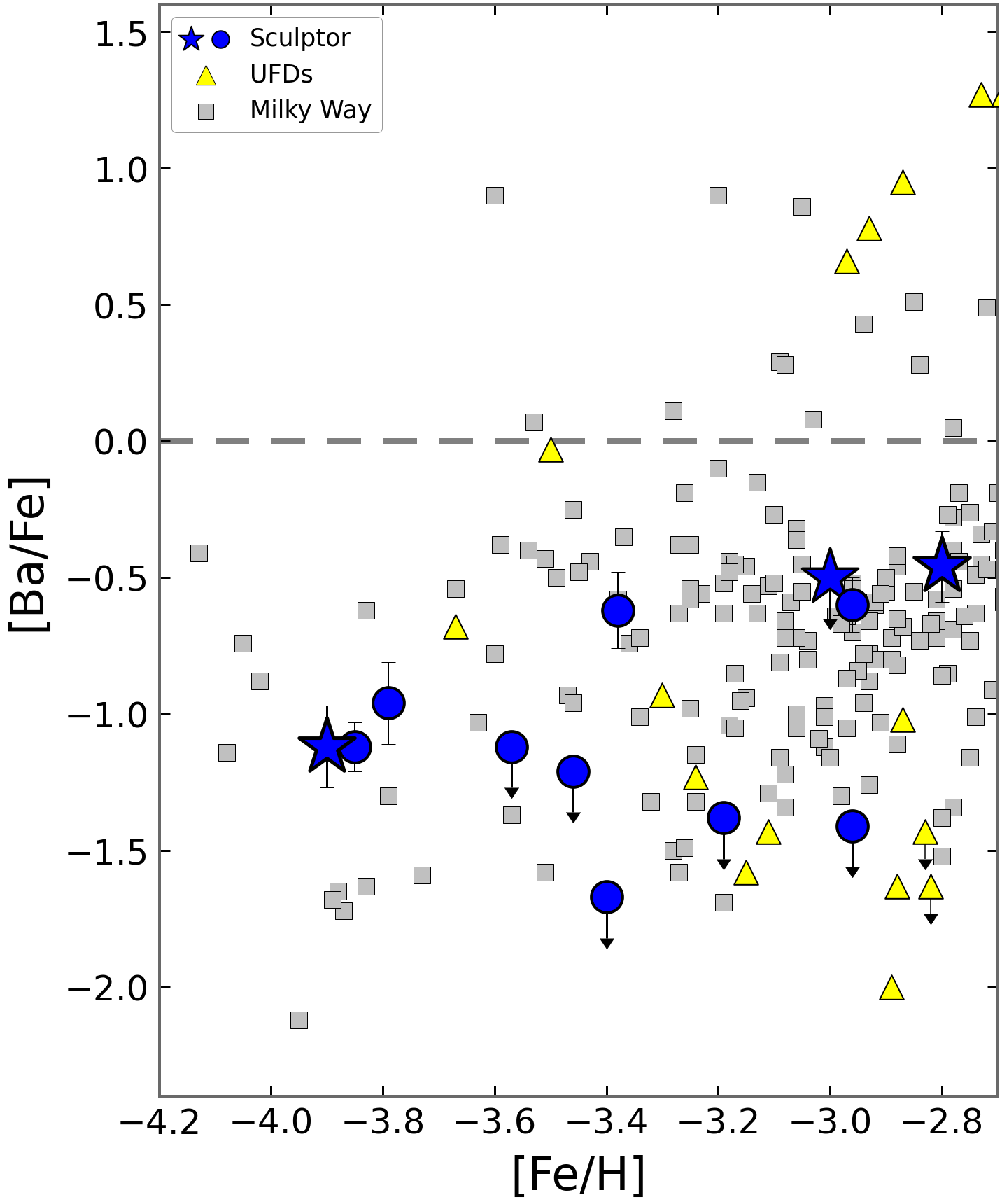}
\hspace{0.2cm}
\includegraphics[width=0.32\linewidth]{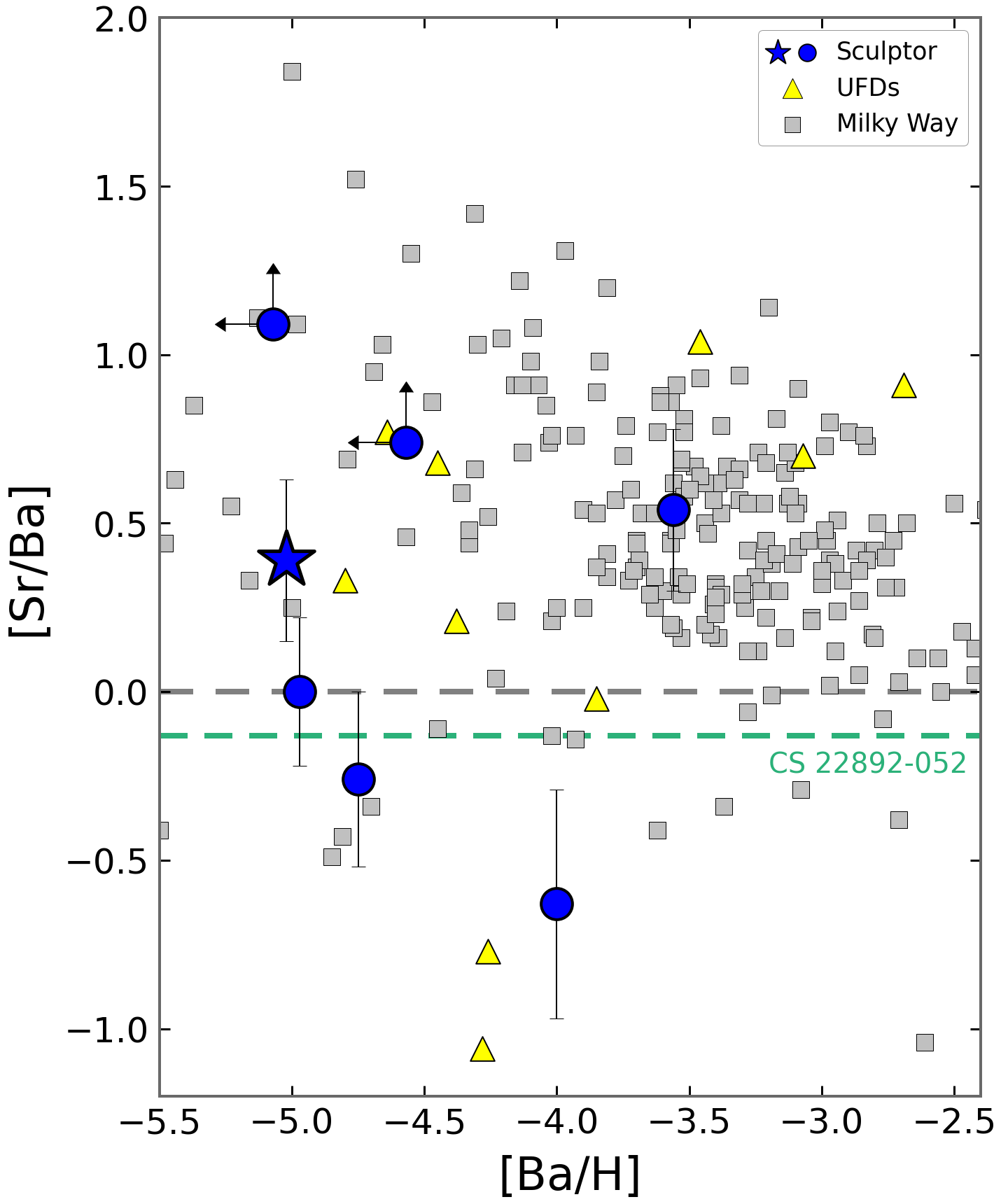}
\caption{Abundance ratios of the neutron-capture elements in Sculptor (blue), the Milky Way (gray) and UFDs (yellow): [Sr/Fe] (left) and  [Ba/Fe] (middle) as a function of [Fe/H], and [Sr/Ba] (right) as a function of [Ba/H]. For reference, the [Sr/Ba] value of the r-process enhanced star CS~22892-052 \citep{Sneden03} is noted with a green dashed line. Star symbols represent Sculptor stars analysed here using new spectra, while blue circles are stars reanalysed on archival spectra. \textit{Milky Way ref.:} \citet{Roederer14}. UFD ref. are listed in the beginning of Sec.~\ref{sec:results}.}
\label{fig:ncapture}
\end{figure*}

\subsection{The $\alpha$-elements}
The $\alpha$-elements Mg, Si, and Ca, along with the often grouped-together Ti, are shown relative to Fe in Fig.~\ref{fig:alpha}. In the elements Mg, and Ca there is a decreasing trend of [Mg,Ca/Fe] towards lower [Fe/H]. This was already identified in the literature compilation of these Sculptor stars \citep{Skuladottir21}, where low Mg and/or low Ca at $\rm [Fe/H]<-3.5$ were previously noted in Sculptor stars by \citet{Tafelmeyer10,Frebel10,Jablonka15}; and \citet{Simon15}. Here we confirm that this is not a result of inhomogenities between different works. This trend is not seen in the Milky Way halo, nor in the UFDs. However, data in other dSph galaxies is lacking at these low $\rm [Fe/H]<-3.5$, hence it is unclear whether this trend is typical for dSph galaxies or if it is unique to Sculptor. 

Our analysis also suggests a lower Si in Sculptor compared to the Milky Way halo and UFDs. This was also found in individual stars by \citet{Tafelmeyer10} and \citet{Simon15}, while the results of \citet{Jablonka15} were more in line with the Milky Way. The star AF20549 was found to have very high $\rm [Si/Fe]=0.96\pm0.39$ by \citet{Frebel10}, which is in agreement with the value obtained here, $\rm [Si/Fe]=0.62\pm0.26$, given the large errors in this moderate S/N spectra. We emphasise that Si is measured using only one line, making it more uncertain than the other $\alpha$-elements reported here. 

Contrary to the $\alpha$-elements, Ti does not show any clear evidence of behaving differently in Sculptor and the other galaxies depicted in Fig.~\ref{fig:alpha}. Furthermore there is no signature of a decreasing nor an increasing trend of [Ti/Fe] with [Fe/H]. Therefore, only the traditional $\alpha$-elements, Mg, Si, and Ca, are different in Sculptor at the lowest $\rm[Fe/H]<-3.5$, compared to the Milky Way halo and the UFDs.

\subsection{Iron-peak elements}
Fig.~\ref{fig:ironpeak} shows the [X/Fe] for the iron-peak elements Sc, Cr, Mn, Co, Ni, and Zn as a function of [Fe/H]. The abundances of [Sc/Fe] in Sculptor are in general agreement with the Milky Way halo and UFDs, especially at the lowest $\rm[Fe/H]<-3.5$, while the higher values of $\rm [Sc/Fe]>+0.2$ are missing in Sculptor while observed in the other galaxies at $\rm [Fe/H]>-3.3$. Similarly to Na, Al, and Sc, [Cr/Fe] in Sculptor also follows the lower envelope of the Milky Way, being consistent with the UFDs. Furthermore, [Co/Fe] is lower in Sculptor compared to the other galaxies. 

On the other hand, the Mn in Sculptor is in very good agreement with both the Milky Way halo and UFDs, showing a very large scatter in [Mn/Fe] and typically subsolar values (in LTE). We note that the three most metal-poor stars in Sculptor generally agree remarkably well in their abundance patterns (Figs~\ref{fig:NaAl}-\ref{fig:ironpeak}), however, in [Mn/Fe] there is significant differences in their values. Furthermore, for these three stars, the [Mn/Fe] is anti-correlated with their \teff\ (see Fig.~\ref{fig:mntriplet}). We note that the abundances presented here, and shown in Fig.~\ref{fig:ironpeak}, have not been corrected for NLTE effects which are predicted to be around $\rm \Delta [Mn/H]_\text{NLTE}\approx+0.5$ in \citet{Bergemann19}, or even as high as $\rm \Delta [Mn/H]_\text{NLTE}\approx+1$ in some works \citep{Bergemann08}. The complete NLTE abundances therefore might reduce the differences. However, the available work on NLTE does not predict such sharp differences in the \teff\ range presented here, therefore the \teff-[Mn/Fe] anti-correlation is plausibly only a coincidence based on low number statistics. 

Finally, both Ni and Zn Sculptor is in very good agreement with the Milky Way halo evolution as well as the UFDs. In particular, Ni, which is easier to measure and thus has less random errors, shows roughly $\rm[Ni/Fe]\approx0$ in Sculptor, as well as the Milky Way and the smaller UFDs. Similar to the Milky Way halo \citep[e.g.][]{Cayrel04, Lombardo22}, Sculptor has increasing [Zn/Fe] ratios toward lower [Fe/H]. Very few Zn measurements are available in UFDs, but in general they are consistent with the larger galaxies. 

Overall we find less differences between Sculptor and the other galaxies in the iron-peak elements, compared to the lighter elements presented in Figs~\ref{fig:C}-\ref{fig:alpha}. Given the uncertainties and possible systematic errors we conclude that the iron-peak elements in Sculptor have a similar behaviour as the other galaxies, albeit not always identical.

\subsection{Neutron-capture elements}

The neutron-capture elements Sr and Ba, relative to Fe, are shown in Fig.~\ref{fig:ncapture}. We note that the four UFD stars with $\rm[Ba/Fe]>0$ and $\rm [Sr/Fe]>0$ belong to $r$-process enhanced Reticulum~II \citep{Ji16c,Ji16d,Roederer16}. Excluding these outliers, the [Sr/Fe] and [Ba/Fe] ratios in Sculptor are in good agreement with UFDs, as well as the Milky Way halo. At the earliest stages in these galaxies, AGB stars had not yet become a dominant source of these elements, leading to extremely low, sub-solar ratios, relative to Fe. We note that the scatter in our [Sr/Fe] and [Ba/Fe] measurements is significant, providing clear evidence of inhomogeneous mixing in these elements at the earliest times in Sculptor. A compatible result (albeit with fewer data points) was found by \citet{Reichert20}, as well as \citet{Mashonkina16}. 

The right panel of Fig.~\ref{fig:ncapture} shows the [Sr/Ba] ratio. Again, Sculptor is in very good agreement with the Milky Way halo and the UFDs at low $\rm[Ba/H]<-3$. As a reference, we denote the [Sr/Ba] of the $r$-process enhanced star CS~22892-052 \citep{Sneden03}. However, very r-process enriched stars, with $\rm [Eu/Fe]>+1$, cover a more than a dex spread in $\rm -0.7\lesssim[Sr/Ba]\lesssim+0.7$ \citep{Holmbeck20}. Similarly, very s-process enriched CEMP-s stars, cover a similar range in [Sr/Ba] \citep[e.g.][]{Hansen19,Goswami21}. Traditionally a specific weak $r$- and/or $s$-processes \citep[e.g.][]{Travaglio04,ArconesMontes11} are provoked to explain the high [Sr/Ba] ratios observed in some halo stars \citep[e.g.][]{Honda04}, in dwarf galaxies \citep[e.g.][]{Skuladottir15a,Susmitha17,Spite18}, and in one $\omega$~Centauri star \citep{Yong17}. The abundance pattern of such stars have been successfully explained, e.g. with fast rotating, low-metallicity massive stars \citep[][]{Cescutti13,Choplin17,Choplin18,Yong17,Skuladottir20}, but the origin of these stars is still being debated. Based on the observational scatter in [Sr/Ba] in both s-process enriched stars as well as r-process rich, it seems likely that the [Sr/Ba] ratio is sensitive to the physical conditions on each site, as well as possibly Sr having contribution from several processes.

\begin{figure}
\centering
\includegraphics[width=0.95\linewidth]{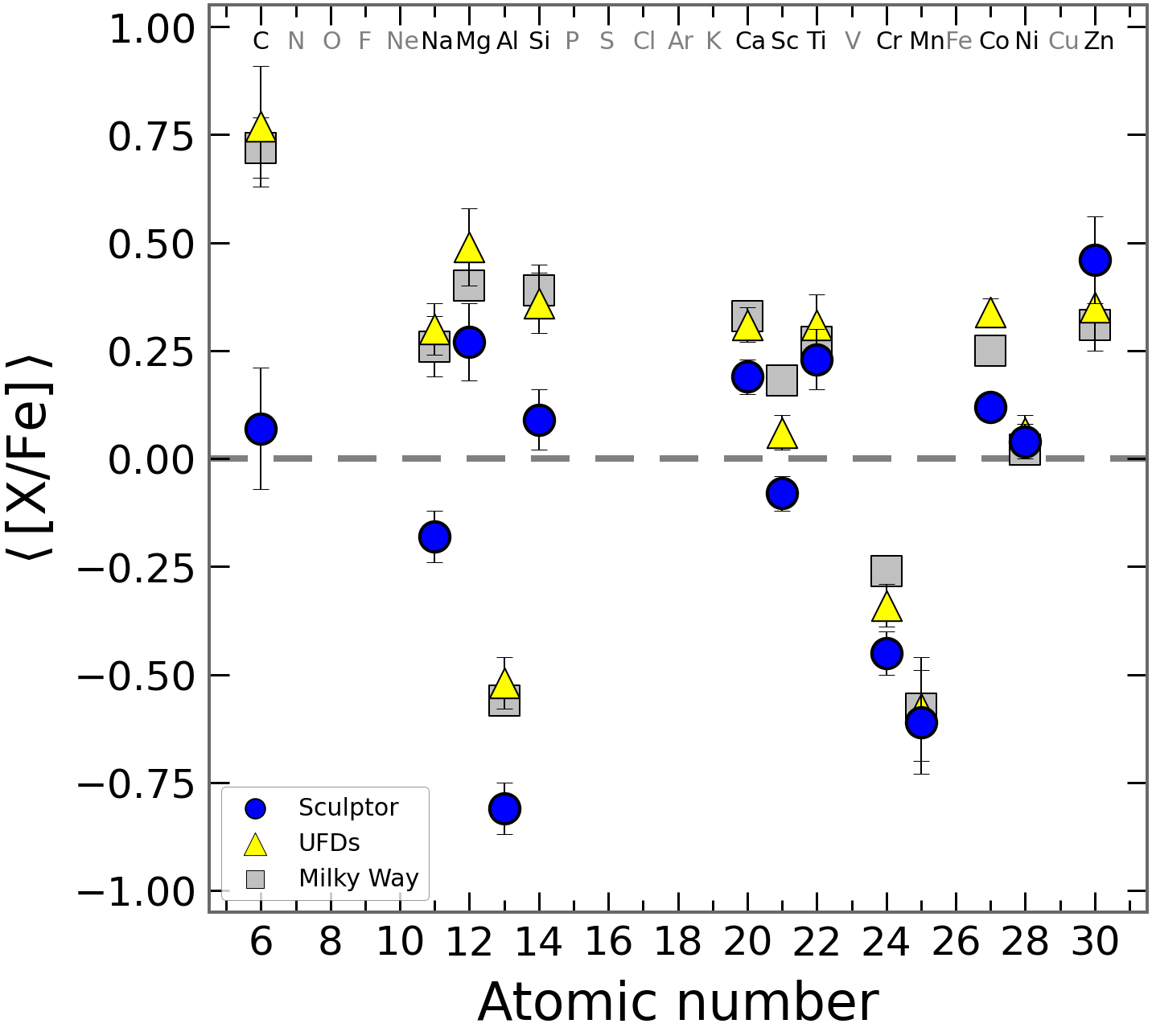}
\caption{The average abundance pattern at $\rm -4\leq[Fe/H]\leq-2.7$ in Sculptor (blue), the Milky Way (gray), and UFDs (yellow), with the error of the mean, $\sigma$/$\sqrt{N-1}$. Informative upper limits (lower than the mean value) are included. Furthermore, all C measurements and upper limits in Sculptor are included. References are the same as in Figs \ref{fig:C}-\ref{fig:ironpeak}.}
%from metov.py
\label{fig:pattfig}
\end{figure}

\begin{figure}
\centering
\includegraphics[width=\linewidth]{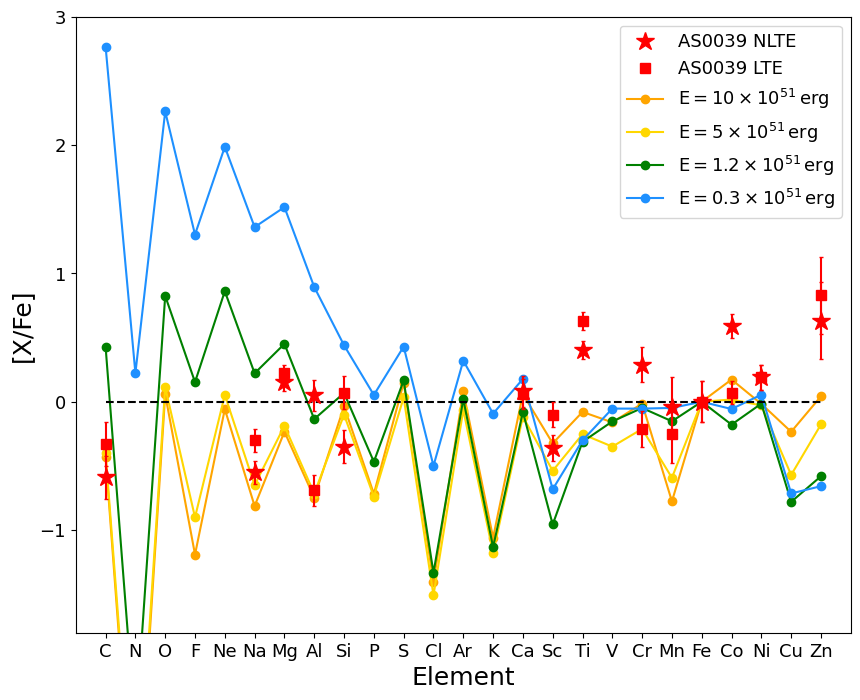}
\caption{Comparison between the measured chemical abundance pattern, [X/Fe], of the star AS0039 (red) in LTE (squares) and NLTE (stars) with model predictions (lines) for the pure descendants of a Pop~III $20\,\mathrm{M_{\odot}}$ star with fixed mixing $f_{mix}=0.063$. Different colors denote different SN explosion energies for Pop~III stars: $0.3 \times 10^{51}$\,erg (blue), $1.2 \times 10^{51}$\,erg (green), $5 \times 10^{51}$\,erg (yellow), and $10 \times 10^{51}$\,erg (orange).}
%from metov.py
\label{fig:abu_patt_ire}
\end{figure}

\subsection{The general abundance pattern} \label{sec:genabu}

Fig.~\ref{fig:pattfig} shows the average abundance pattern from the data shown in Figs~\ref{fig:C}-\ref{fig:ironpeak}, at the metallicity range of the Sculptor data, $\rm -4<[Fe/H]<-2.7$, with the error of the mean, $\sigma$/$\sqrt{N-1}$. Upper limits in Sculptor and the UFDs that are informative, i.e. lower than the mean [X/Fe], are included as measurements. All upper limits in [C/Fe] are included for the Sculptor average. Despite this very conservative estimate of $\rm \langle[C/Fe]\rangle_\textsl{Scl}$, it is clear that Sculptor is remarkably devoid of C compared to the Milky Way halo and the UFDs (see also Fig.~\ref{fig:C}).

Furthermore, the lighter elements show a similar trend, where in particular Na, Al, and Si have significantly lower values in Sculptor compared to other galaxies. Less significantly, [Mg/Fe] and [Ca/Fe] in Sculptor are lower than in the Milky Way and UFDs. We note that Mg and Ca are consistent with the other systems at higher $\rm[Fe/H]\gtrsim-3.5$, and only deviate at the lowest [Fe/H]. However, when looking at elements from Ti and heavier in Fig.~\ref{fig:pattfig} it is clear that the abundance pattern in Sculptor converges with the Milky Way halo and the UFDs, and the different galaxies are generally in very good agreement.

The peculiarities of the Sculptor average abundance pattern are likely closely associated with the galaxy's dearth of CEMP-no stars (Sec.~\ref{sec:carbon}; e.g.~\citealt{Skuladottir15a}). Indeed, CEMP-no stars - especially at low [Fe/H] - have shown an enhancement of the lighter elements ($Z\lesssim20$), while the heavier elements are more in line with C-normal stars in the Milky Way halo (e.g.~\citealt{Norris13}). This is clear evidence that low-energy SN, the progenitors of CEMP-no stars (e.g.~\citealt{Iwamoto05}; Rossi et al. 2023; Vanni et al. 2023), have not had as great an impact on Sculptor as the Milky Way halo and the UFDs. Ultimately, this raises the question from where the earliest chemical enrichment in Sculptor arises.

\section{Primordial Hypernovae in Sculptor} \label{sec:models}

To interpret the chemical abundances of the Sculptor stars we use the parametric model for early chemical enrichment presented in \citet{Salvadori19} to study the signature of very massive Pop~III stars and then extended in Vanni et al. (2023 in press; and 2023 in prep.) to uncover the imprint of Pop~III stars with different mass and explosion energy. This broad and general model studies the chemical abundances in the primordial galaxies after the pollution of a single Pop~III star with mass, $M_* \in [10; 1000] \,\mathrm{M_{\odot}}$, SN explosion energy, $E_\textsl{SN} \in [0.3; 100] \times 10^{51}$ erg and mixing, $f_{mix} \in [0; 0.25]$. We exclude, for this work, stars with $M_* >100 \: \mathrm{M_{\odot}}$ (and $E_\textsl{SN} > 10 \times 10^{51}$ erg) and use for the remaining ones the yields from \citet{Hegerwoosley10}. Moreover, the model explores how these abundances change with the contamination of normal Pop~II stars at different levels. The unknowns on the early phases of the stellar formation and evolution are wrapped in three free parameters: the star formation efficiency of Pop~III stars, the dilution factor of the metals and the fraction of metals coming from Pop~III stars with respect to the total. See Vanni et al. (2023, in press) and \citet{Salvadori19} for further details and a comprehensive explanation of the model properties.

\subsection{The progenitor of AS0039}

The star AS0039 is the most metal-poor star known in any external galaxy. The abundance pattern of this star is in general agreement with other two Sculptor stars at similar $\rm[Fe/H]_{LTE}<-3.7$ (see Fig.~\ref{fig:C}-\ref{fig:ncapture}), and due to its brightness (see Fig.~\ref{fig:radecrgb}, and Table~\ref{tab:stellsample}) its S/N is higher (see e.g.~Fig.~\ref{fig:mntriplet}). Thus in the following, we use AS0039 as representative of the earliest chemical enrichment in Sculptor and use the NLTE abundances provided in Table~\ref{tab:as0039}. 

In Fig. \ref{fig:abu_patt_ire} we compare the abundances of AS0039 computed in both LTE (square points) and NLTE (star points), with the abundances predicted by our model for the pure (100\%) descendants of Pop~III stars with fixed mass, $20\,\mathrm{M_{\odot}}$, and mixing, $f_{mix}=0.063$, for four different SN explosion energies \citep{Hegerwoosley10}. 

First, we notice that the abundances of AS0039 better agree with the descendants of high-energy, $E_\textsl{SN} \geq 5 \times 10^{51}$ erg, Pop~III SNe (yellow and orange points), that are the only ones for which the model predicts [C/Fe] $< 0$. Moreover, our models predict very similar abundances for the descendants of low- and high-energy SNe for the elements heavier than Sc, while they are very different for the lighter metals, most of all C. It suggests that the abundances of the elements lighter than Sc might be the key to understand the pollution history of AS0039 and the other environments presented in Fig.~\ref{fig:pattfig}: the Milky Way stellar halo, UFDs and Sculptor. Indeed, the abundance ratios presented in the Figs.~\ref{fig:pattfig} and \ref{fig:abu_patt_ire}, if compared, suggest that, on average, Sculptor has been polluted by more energetic primordial SNe with respect to the Milky Way halo and the UFDs, which, on the other side, might have been more dominated by the ejecta of low-energy Pop~III SNe (\citet{Rossi2023}; and in prep).

\begin{figure}
\centering
\includegraphics[width=\linewidth]{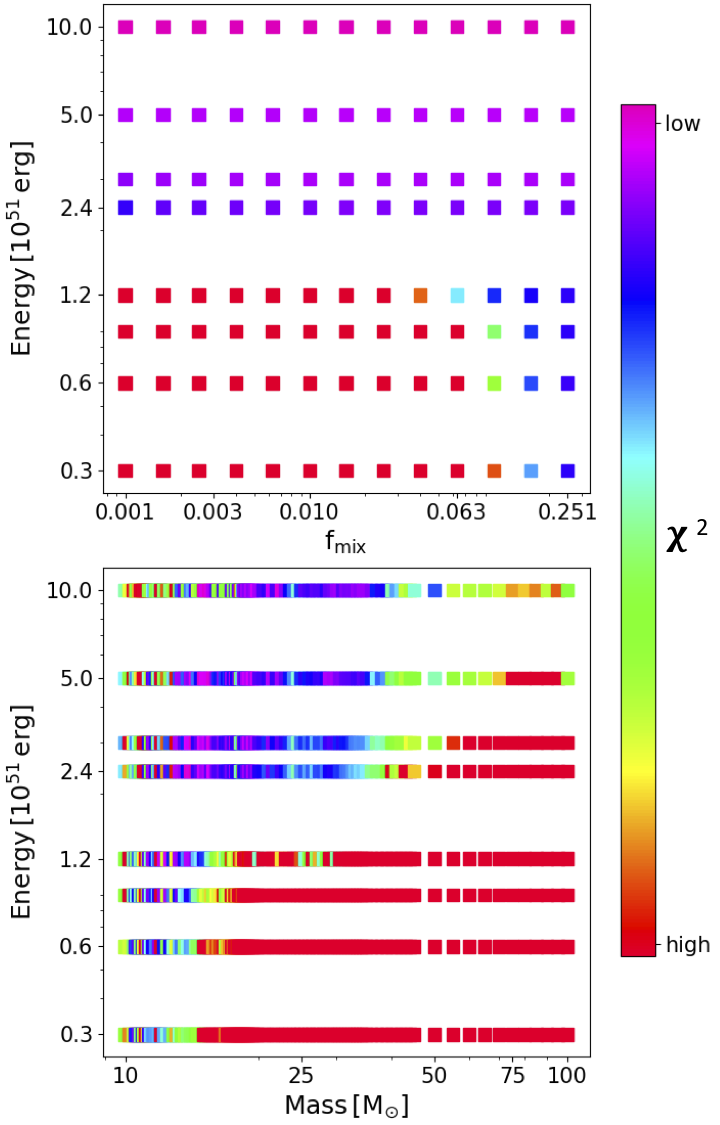}
\caption{Comparison of the observed abundances of AS0039 (NLTE) to theoretical predictions for pure descendants of Pop~III stars, varying their SN explosion energy, mixed fraction and initial mass. In the upper panel we fixed the initial mass to $\mathrm{20\,M_{\odot}}$, in the lower panel the mixed fraction to $\mathrm{f_{mix}=0.063}$. Best fits are magenta, while the worst fits, with $\chi^2\geq5$ (i.e.~$\geq5$ times the minimum), are shown with red color. 
}
\label{fig:chi_square_ire}
\end{figure}

\begin{figure*}
\centering
\includegraphics[width=0.6\linewidth]{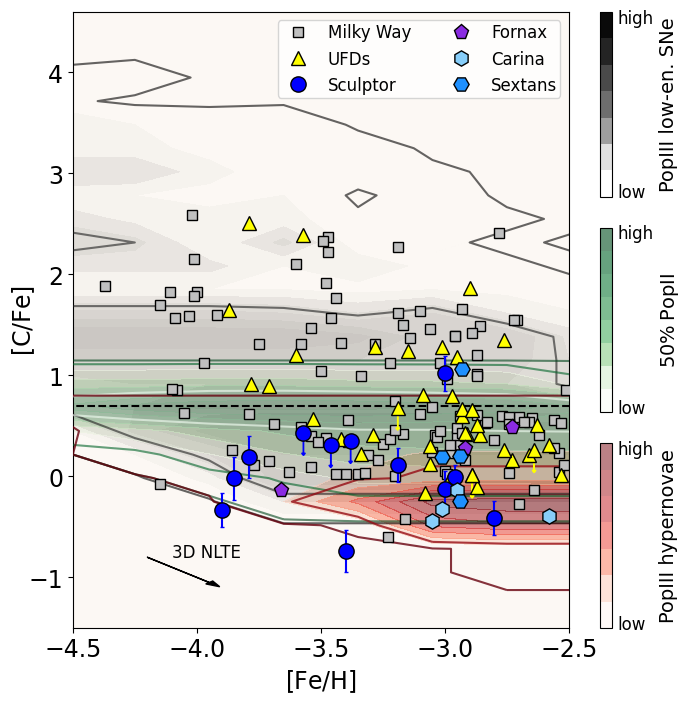}
\caption{Density maps of the [C/Fe] with respect to [Fe/H] as predicted by our model for the descendants of Pop~III stars (shaded areas) and measured abundances (LTE) of stars in Sculptor (blue circles), Sextans (medium-blue hexagons), Carina (light-blue hexagons) and Fornax (purple pentagons), the UFDs (yellow triangles), and the Milky Way (grey squares). All measured [C/Fe] abundances have been corrected for internal mixing \citep{Placco14}. Black arrow shows the expected size and direction of how measured stellar abundances will change because of NLTE effects in [Fe/H]. The shaded areas are colored depending on the type of stars that polluted the descendants: 100\% Pop~III low-energy SNe (grey, $E_\textsl{SN} < 1.5 \times 10^{51}$ erg), 100\% Pop~III hypernovae (red, $E_\textsl{SN} > 5.0 \times 10^{51}$ erg), 50\% low-energy Pop~III and 50\% core-collapse Pop~II SNe (green). The densest areas (and most probable abundances) are coloured with darker shades. In all cases $f_{mix}=0.063$.  Ref. for the Milky Way and UFDs are listed in Sec.~\ref{sec:results}. Ref for dSph galaxies: \citet{Tafelmeyer10}; \citet{Lucchesi2020}; Lucchesi et al. in prep.}
\label{fig:density_map_ire}
\end{figure*}

%\subsection{Chi-square}

Given the good agreement between AS0039 and the descendants of high-energy Pop~III SNe for a given mass and mixing, we directly compare its abundances with theoretical models varying the mass, the SN explosion energy and the mixing of the Pop~III progenitor star. We compute the $\chi^2$ between the chemical abundances of AS0039 and the ones predicted by 100\% Pop~III descendants as follows:
\begin{equation}
    \rm \chi^2_\circ=\sum^N_{i=1} \dfrac{\left([X_i/Fe]_{obs}-[X_i/Fe]_{th}\right)^2}{ \rm \left(\sigma^2_{[X_i/Fe]_{obs}}+\sigma^2_{[X_i/Fe]_{th}}\right)},
\end{equation}
where $\rm [X_i/Fe]_{obs, th}$ and $\rm \sigma^2_{[X_i/Fe]_{obs, th}}$ are the measured and theoretical abundance ratios and standard deviations. In order to account for the uncertainties on Pop~III stellar yields, quantified by \citet{Nomoto2013} and \citet{Hartwig2018} in $\rm \sigma_{[X_i/Fe]_{th}} \sim 0.3-1.0~dex$, we assume $\rm \sigma^2_{[X_i/Fe]_{th}}=0.25$. Finally, we define scaled $\rm \chi^2=\chi^2_\circ/$min($\chi^2_\circ$).

The best fit suggests that AS0039 was primarily enriched by a Pop~III hypernovae, $E_\textsl{SN} = 10 \times 10^{51}$ erg, with $M_* \approx 20\,\mathrm{M_{\odot}}$, regardless of the mixing. This results is in agreement with the best fit computed with the online abundance-fitting tool Starfit\footnote{https://starfit.org/} \citep[see][]{Hegerwoosley10}. However, there are degeneracies in the models and, taking into account uncertainties of the chemical abundances, there can be several models that produce satisfactory fits. Therefore it is important to explore the parameter space, to ensure that the best fit does not represent a single solution, but is representative of the best solution space.

The results of the $\chi^2$ analysis are presented in Fig. \ref{fig:chi_square_ire}, where we show the goodness of fit with different colors when varying the mass, the mixing and the SN explosion energy of the Pop~III progenitor. To simplify, we provide examples with fixed $M_*=20\,\mathrm{M_{\odot}}$, in the top panel, and fixed $f_{mix}=0.063$, in the bottom panel. Regardless of the mixing, for a 20\,$\mathrm{M_{\odot}}$ Pop~III progenitor, the descendants of Pop~III hypernovae ($E_{SN} \geq 5 \times 10^{51}$\,erg) have the minimum $\chi^2$, while the worst agreement is with the descendants of Pop~III SNe with $E_{SN} \leq 1.2 \times 10^{51}$ erg. For fixed $f_{mix}$, it is evident that the number of descendants with a good agreement with AS0039 drastically decreases for the lowest-energy Pop~III SN progenitors. Indeed, we find the minimum $\chi^2$ for the descendants of high-energy SNe ($E_{SN} \geq 2.4 \times 10^{51}$ erg) with masses between 12 $\mathrm{M_{\odot}}$ and 40 $\mathrm{M_{\odot}}$. 

The best fit to the abundance pattern of AS0039, as well as exploration of the parameter space (Figs.\ref{fig:abu_patt_ire}-\ref{fig:chi_square_ire}), strongly suggest this star was primarily enriched by a primordial Pop~III hypernova, with $M\approx20$\,M$_\odot$. Since AS0039 has a similar abundance pattern to other low $\rm [Fe/H]_{LTE}<-3.7$ stars in the galaxy, this points to an earliest chemical enrichment in Sculptor that was dominated by high-energy SNe. %{\bf Indeed, if the same NLTE corrections are adopted for these stars as for AS0039, MT00750 is best fitted with the yields of a 15\,$\mathrm{M_{\odot}}$ hypernovae, $E_{SN} \geq 10 \times 10^{51}$\,erg, and PJ03111 is best fit with a high energy SN, $E_{SN} \geq 2.4 \times 10^{51}$\,erg, with a similar mass.}
Assuming the same NLTE corrections derived for AS0039, we find that also the two stars MT00750 and PJ03111 are best fitted by the descendants of high-energy Pop~III SNe. In particular, MT00750 agrees with the descendant of a $15 \:\mathrm{M_{\odot}}$ hypernova ($E_{SN} = 10 \times 10^{51}$ erg) without dependence on $f_{mix}$, while PJ03111 with the descendant of a $14.4 \:\mathrm{M_{\odot}}$ high-energy SN ($E_{SN} = 2.4 \times 10^{51}$ erg) with $f_{mix}=0.0251$.

\subsection{The imprints of high-energy SN in Sculptor}

In Fig.~\ref{fig:density_map_ire} we compare the results of our model (Vanni et al. 2023, in press) for [C/Fe] as a function of [Fe/H] for observational data in the dSph galaxies: Sculptor, Fornax, Carina and Sextans; along with those observed in the Milky Way and UFDs. We note that literature data are typically only provided in LTE, however, NLTE affects in iron in metal-poor giants are expected to be on the order of $\rm \Delta[Fe/H]\approx+0.25$ (see Table~\ref{tab:as0039}), and will effect [Fe/H] as well as [C/Fe] (shown with arrow in Fig.~\ref{fig:density_map_ire}). The [C/Fe] will be even lower when taking 3D effects into account, affecting C measurements from molecular lines \citep[][]{NorrisYong19}.

We show the [C/Fe] abundances predicted by our model for the descendants of pure low energy Pop~III SN (gray), and pure hypernovae descendants (red), i.e. $100\%$ metals from these different types of Pop~III SNe. Furthermore, we show an intermediate case, where the descendants have been polluted only at a 50\% level by low-energy Pop~III SNe and the remaining 50\% metals are provided by normal Pop~II SNe (green area). 
The density diagrams are created exploring all the possible initial masses of Pop~III stars (i.e. $10-100\,M_{\odot}$), with a mixing of $f_{mix}=0.063$. The contribution of Pop~II stars is, on the other hand, integrated over a normal Larson IMF
\citep[see][]{Salvadori19}. 

The majority of the stars pertaining to the dSph galaxies are carbon normal (the maximum is $\rm[C/Fe]\approx +1$), while the MW and UFDs have stars with $\rm[C/Fe]\geq +2.0$. The high values of [C/Fe] in these two latter environments can be explained only with the pure descendants of low-energy Pop~III SNe, with $E_\textsl{SN} < 1.5 \times 10^{51}$ erg, which expel low amounts of iron (see Vanni et al. 2023; and submitted to MNRAS). The stars with lower [C/Fe] values, agree better with the pollution of both low-energy Pop~III SNe and Pop~II stars. However, when taking into account the predicted NLTE effects, the [C/Fe] values of most stars in Sculptor and Carina are too low to be explained with these two cases. Indeed, their very low [C/Fe] value can only be explained by a strong imprint by Pop~III hypernovae.

\section{Conclusions}

We present an analysis of a newly obtained HR UVES spectrum for AS0039 in the Sculpor dSph galaxy, providing both LTE and NLTE abundances. This star is confirmed to be the most-metal-poor star in any external galaxy, with $\rm[Fe/H]_{LTE}=-3.90$ and $\rm[C/Fe]_{LTE}=-0.33$ (corrected for internal mixing). To avoid systematic errors, we re-analyse all the archival spectra available to us at $\rm[Fe/H]\leq-3$ in Sculptor, and provide an analysis for two stars published for the first time in this work: UHAL004 at $\rm[Fe/H]=-2.8$ and a CEMP-no star, DR20080, discovered here at $\rm[Fe/H]=-3$.

Out of the 11 Sculptor stars with carbon measurements at $\rm[Fe/H]\leq-3$, only one is a CEMP-no star, giving a fraction of $f_\textsl{CEMP}^\textsl{Scl}=9^{+11}_{-8}\%$, at odds with fractions measured in the Milky Way and UFDs, $f_\textsl{CEMP}\approx40\%$ \citep{Placco14,Ji20}. This is a further confirmation that Sculptor is remarkably devoid of CEMP-no stars, as noted by previous works \citep{Starkenburg13,Skuladottir15a,Skuladottir21}.

Here, we show clearly that these peculiarities in Sculptor reach beyond carbon. Compared to the Milky Way and UFDs, the metal-poor Sculptor stars have lower [X/Fe] ratios in light elements (e.g. Na, Al, Mg, Si). On the other hand, heavier elements, from Ti and beyond, are generally in good agreement with chemical abundances both in the Milky Way halo and UFDs. This unique abundance pattern in Sculptor is consistent with enrichment from Pop~III stars with high explosion energy, $E_\textsl{SN} > 2 \times 10^{51}$\,erg, which are predicted to produce low ratios of [X/Fe] for lighter elements, while heavier elemental ratios ([X/Fe], $Z\gtrsim20$) are less sensitive to explosion energy \citep{HegerWoosley02}. The average abundance pattern of the Milky Way and UFDs is remarkably similar, while being clearly different from that of Sculptor. This result is somewhat at odds with the prediction that the Milky Way halo is mostly made up by massive dSph galaxies \citep[e.g.][]{Deason16}. However, it is unclear how the earliest chemical enrichment of those Milky Way progenitors differentiated to that of the present day satellite dSph galaxies.

When comparing our Sculptor results with theoretical models, the (NLTE) abundance pattern of the most metal-poor star in our sample, AS0039, is best explained by it being mainly enriched by a zero-metallicity hypernovae, $E_\textsl{SN} =10 \times 10^{51}$ erg, with a mass $M=20$\,M$_\odot$ in good agreement with previous analyses of the discovery X-Shooter spectrum \citep{Skuladottir21}. This result is robust, and obtained regardless of the adopted mixing.

The [C/Fe] abundances measured in Sculptor were compared to our general parametric model, further confirming this conclusion. 
The very low [C/Fe] abundances suggest that the Sculptor dSph galaxy hosted at least one Pop~III hypernova, which dominated its earliest enrichment, overshadowing the impact of lower energy SNe. We emphasise that Pop~III faint SN are still likely to have occurred in the galaxy (and/or in its progenitors) since they are needed to explain the existence of CEMP-no stars \citep[see also][]{Salvadori15}, which have been detected at $\rm[Fe/H]=-3$ (DR20080), and $\rm[Fe/H]=-2$ \citep{Skuladottir15a}.
%The very low [C/Fe] abundances suggest that the Sculptor dSph galaxy experienced a Pop~III hypernova, dominating the early enrichment, overshadowing the impact of lower energy SNe. We emphasise that Pop~III faint SN are still likely to have occurred in the galaxy (and/or in its progenitors), as seen by the detected CEMP-no stars at $\rm[Fe/H]=-3$ (DR20080), and $\rm[Fe/H]=-2$ \citep{Skuladottir15a}.

Our results show that Sculptor is massive enough to be able to retain the products of such energetic events. This suggests that dSph galaxies are key objects to study the full parameter space of the Pop~III stars, i.e. their mass distribution and the energy distribution of their SNe. This is further supported by available data in other dSph galaxies, which have low [C/Fe] and relatively few CEMP-no stars, compared to the Milky Way and UFDs. 
%Our results show that Sculptor is large enough to be able to retain the products of such energetic events, while UFDs might have been completely quenched if experiencing such an event. This suggests that dSph galaxies are necessary to study the full parameter space of the first stars. This is further supported by available data in other dSph galaxies, which have low [C/Fe] and relatively few CEMP-no stars, compared to the Milky Way and UFDs. 

The upcoming large spectroscopic surveys will provide more data on dSph galaxies, in particular \textit{4DWARFS: the 4MOST survey of dwarf galaxies and their stellar streams} (Skúladóttir et al. 2023, ESO Messenger, in press) is expected to make a great leap forward. However, to be able to fully exploit the data to reveal the properties of the first stars, sophisticated physically-driven models (e.g. \citealt{Rossi21}; Rossi et al. 2023), targeted at the earliest chemical enrichment in dSph galaxies are crucially needed. By combining data with models, the Sculptor dwarf spheroidal galaxy - and other similar galaxies - might be our best chance at understanding the energy distribution of the first supernovae in the Universe.

\vspace{0.5cm}

\begin{acknowledgements}
The authors thank J.~Simon and A.~Frebel for readily sharing their data.
This project has received funding from the European Research Council (ERC) under the European Union’s Horizon 2020 research and innovation programme (grant agreement No. 804240).
This work has made use of data from the European Space Agency (ESA) mission
{\it Gaia} (\url{https://www.cosmos.esa.int/gaia}), processed by the {\it Gaia}
Data Processing and Analysis Consortium (DPAC,
\url{https://www.cosmos.esa.int/web/gaia/dpac/consortium}). Funding for the DPAC
has been provided by national institutions, in particular the institutions
participating in the {\it Gaia} Multilateral Agreement.
\end{acknowledgements}

\bibliographystyle{aa}
\bibliography{heimildir}

\clearpage


\end{document}

%% file: tables/newdata.tex
\begin{table}
\caption{Observational log for the new spectra, analysed here for the first time.}

\label{tab:newdata}
\centering
\scriptsize
\tabcolsep=0.07cm
\renewcommand{\arraystretch}{1.3}

\begin{tabular}{llccrc}
\hline\hline

Star & ESO facility & Exp time & $\lambda$  & Resolution & S/N \\
&&[h] & [nm]&& [pxl$^{-1}$]\\
\hline

AS00039 & UVES (blue)& 12 & 326-454 & 20\,000 & 31@444nm\\
 & UVES (red) & 12 & 458-668 & 40\,000 & 43@666nm\\
\hline
DR20080 & X-Shooter (UVB) & 1 & 300-550 & 5\,400 & 12@444nm\\
 & X-Shooter (VIS) & 1 & 550-1020 & 8\,900 & 18@600nm\\
\hline
UHAL004 & FLAMES/GIRAFFE & 4 & 397-427 & 6\,000 & 26@444nm\\
& FLAMES/UVES & 20 & 480-680 & 47\,000 & 32@600nm\\

\hline\hline
\end{tabular}
\end{table}

%% file: tables/stellsample.tex
\begin{table}
\caption{Positions, magnitudes and atmospheric parameters of the stellar sample, including references to literature work (in parenthesis when the spectrum is different from that analysed here). Typical random errors on the stellar parameters are: $\Delta\teff=86\,$K; $\Delta\logg=0.14$; and $\Delta\vt=0.1\,\kms$.
}
\label{tab:stellsample}
\scriptsize
\tabcolsep=0.08cm
\renewcommand{\arraystretch}{1.2}

\begin{center}
\begin{tabular}{lcccccccc}

\hline\hline
Star	&	RA	&	Dec.	&	G	&	\teff	&	\logg	&	\vt	&	Ref.	\\
	&	hh:mm:ss	&	hh:mm:ss	&	mag	&	[K]	&		&	\kms	&		\\
\hline															
HR sample	&		&		&		&		&		&		&		\\
\hline															
AS00039	&	00:58:45.60	&$-$	33:42:21.8	&	16.932	&	4372	&	0.76	&	2.0	&	(1)	\\
AF20549	&	01:00:47.84	&$-$	33:41:03.2	&	17.940	&	4758	&	1.37	&	1.8	&	2,3	\\
JS14296	&	00:59:38.76	&$-$	33:46:14.6	&	18.783	&	4929	&	1.79	&	1.7	&	3	\\
JS66402	&	01:00:00.41	&$-$	33:29:15.5	&	18.774	&	4985	&	1.81	&	1.7	&	3	\\
MT00749	&	01:00:05.02	&$-$	33:61:16.6	&	17.973	&	4797	&	1.41	&	1.8	&	3,4	\\
MT00750$^a$	&	01:00:01.15	&$-$	33:59:21.4	&	18.271	&	4876	&	1.56	&	1.8	&	3,4	\\
PJ00206	&	01:01:26.76	&$-$	33:02:59.8	&	16.687	&	4359	&	0.65	&	2.0	&	5,(6)	\\
PJ03059	&	01:01:22.25	&$-$	33:46:21.9	&	17.538	&	4618	&	1.14	&	1.9	&	5	\\
PJ03111	&	00:57:10.22	&$-$	33:28:35.8	&	17.460	&	4650	&	1.13	&	1.9	&	5,(6)	\\
PJ07402	&	00:57:34.85	&$-$	33:39:45.7	&	17.758	&	4608	&	1.22	&	1.8	&	5,(6)	\\
UHAL004	&	01:01:49.43	&$-$	33:54:10.4	&	16.521	&	4202	&	0.49	&	2.0	&	-	\\
\hline															
LR sample	&		&		&		&		&		&		&		\\
\hline															
DR20080	&	00:57:44.16	&$-$	33:61:14.3	&	17.839	&	4682	&	1.29	&	1.8	&	-	\\
ES03170	&	01:01:47.48	&$-$	33:47:27.6	&	18.385	&	4962	&	1.65	&	1.8	&	6	\\
\hline\hline															
\multicolumn{9}{l}{\fontsize{6.1}{4}\selectfont $^{a)}$ No available spectra of the CH G-band.}\\
\multicolumn{9}{l}{\fontsize{6.1}{4}\selectfont $^{1)}$ \citealt{Skuladottir21}; $^{2)}$ \citealt{Frebel10}; $^{3)}$ \citealt{Simon15}; $^{4)}$ \citealt{Tafelmeyer10};} \\
\multicolumn{9}{l}{\fontsize{6.1}{4}\selectfont   $^{5)}$ \citealt{Jablonka15}; $^{6)}$ \citealt{Starkenburg13}}\\
\end{tabular}
\end{center}

\end{table}

%% file: tables/abundances.tex
\begin{sidewaystable*}
\caption{The chemical abundance measurements (LTE) for the Sculptor stellar sample.}
\label{tab:abundances}
\small
\tabcolsep=0.08cm

\renewcommand{\arraystretch}{1.4}
\centering
\begin{tabular}{lcrcrrrrrrrrrrrrrrrrrrrr}
\hline\hline	
\multicolumn{24}{l}{HR sample}\\
\hline
	&		&	\multicolumn{2}{c}{AS0039}			&	\multicolumn{2}{c}{UHAL004}			&	\multicolumn{2}{c}{AF20549}			&	\multicolumn{2}{c}{JS14296}			&	\multicolumn{2}{c}{JS66402}			&	\multicolumn{2}{c}{MT00749}			&	\multicolumn{2}{c}{MT00750}			&	\multicolumn{2}{c}{PJ00206}			&	\multicolumn{2}{c}{PJ03059}			&	\multicolumn{2}{c}{PJ03111}			&	\multicolumn{2}{c}{PJ07402}			\\

El.	&	$\log\epsilon_\odot$	&	[X/Fe]	&	$\delta_\text{[X/Fe]}$	&	[X/Fe]	&	$\delta_\text{[X/Fe]}$	&	[X/Fe]	&	$\delta_\text{[X/Fe]}$	&	[X/Fe]	&	$\delta_\text{[X/Fe]}$	&	[X/Fe]	&	$\delta_\text{[X/Fe]}$	&	[X/Fe]	&	$\delta_\text{[X/Fe]}$	&	[X/Fe]	&	$\delta_\text{[X/Fe]}$	&	[X/Fe]	&	$\delta_\text{[X/Fe]}$	&	[X/Fe]	&	$\delta_\text{[X/Fe]}$	&	[X/Fe]	&	$\delta_\text{[X/Fe]}$	&	[X/Fe]	&	$\delta_\text{[X/Fe]}$	\\
\hline																																															
Fe	&	7.46	&$	-3.90	$&	0.11	&$	-2.80	$&	0.12	&$	-3.46	$&	0.12	&$	-3.57	$&	0.12	&$	-3.38	$&	0.14	&$	-2.96	$&	0.12	&$	-3.85	$&	0.11	&$	-3.40	$&	0.11	&$	-2.96	$&	0.11	&$	-3.79	$&	0.12	&$	-3.19	$&	0.11	\\
C$_\textsl{obs}$	&	8.46	&$	-0.96	$&	0.20	&$	-1.18	$&	0.20	&$	<-0.15	$&	$-$	&$	<0.36	$&	$-$	&$	<0.27	$&	$-$	&$	-	$&	$-$	&$	-0.21	$&	0.24	&$	-1.46	$&	0.24	&$	-0.75	$&	0.17	&$	-0.48	$&	0.24	&$	-0.57	$&	0.20	\\
C$_\textsl{corr}$	&	8.46	&$	-0.33	$&	0.20	&$	-0.41	$&	0.20	&$	<0.31	$&	$-$	&$	<0.43	$&	$-$	&$	<0.35	$&	$-$	&$	-	$&	$-$	&$	-0.02	$&	0.24	&$	-0.74	$&	0.24	&$	-0.01	$&	0.17	&$	0.19	$&	0.24	&$	0.11	$&	0.20	\\
Na	&	6.22	&$	-0.30	$&	0.07	&$	0.20	$&	0.09	&$	-0.26	$&	0.08	&$	-0.23	$&	0.13	&$	-0.15	$&	0.13	&$	-0.24	$&	0.08	&$	-0.07	$&	0.09	&$	-	$&	$-$	&$	-	$&	$-$	&$	-0.28	$&	0.08	&$	-0.31	$&	0.11	\\
Mg	&	7.55	&$	0.22	$&	0.05	&$	0.54	$&	0.08	&$	0.04	$&	0.24	&$	-0.42	$&	0.17	&$	0.37	$&	0.26	&$	0.55	$&	0.07	&$	0.22	$&	0.09	&$	0.31	$&	0.11	&$	0.54	$&	0.06	&$	0.29	$&	0.15	&$	0.33	$&	0.09	\\
Al	&	6.43	&$	-0.69	$&	0.10	&$	-	$&	$-$	&$	<-1.07	$&	$-$	&$	<-0.76	$&	$-$	&$	<-0.85	$&	$-$	&$	-	$&	$-$	&$	-0.48	$&	0.20	&$	-0.87	$&	0.17	&$	-0.72	$&	0.15	&$	-0.86	$&	0.15	&$	-1.02	$&	0.22	\\
Si	&	7.51	&$	0.07	$&	0.15	&$	-	$&	$-$	&$	0.62	$&	0.27	&$	0.02	$&	0.31	&$	-0.03	$&	0.31	&$	-	$&	$-$	&$	0.00	$&	0.17	&$	-0.08	$&	0.20	&$	0.05	$&	0.18	&$	0.10	$&	0.16	&$	0.09	$&	0.24	\\
Ca	&	6.30	&$	0.06	$&	0.11	&$	0.30	$&	0.09	&$	0.10	$&	0.09	&$	0.25	$&	0.17	&$	0.03	$&	0.18	&$	0.22	$&	0.07	&$	0.09	$&	0.10	&$	0.28	$&	0.08	&$	0.34	$&	0.07	&$	0.02	$&	0.17	&$	0.39	$&	0.06	\\
Sc	&	3.14	&$	-0.10	$&	0.08	&$	-	$&	$-$	&$	0.01	$&	0.12	&$	-0.02	$&	0.18	&$	-0.12	$&	0.14	&$	-	$&	$-$	&$	0.08	$&	0.07	&$	-0.07	$&	0.28	&$	-0.04	$&	0.13	&$	-0.11	$&	0.06	&$	-0.35	$&	0.12	\\
Ti	&	4.97	&$	0.63	$&	0.06	&$	0.41	$&	0.16	&$	0.08	$&	0.08	&$	0.30	$&	0.09	&$	0.11	$&	0.10	&$	0.02	$&	0.07	&$	0.24	$&	0.07	&$	0.03	$&	0.06	&$	0.39	$&	0.07	&$	0.37	$&	0.07	&$	-0.09	$&	0.08	\\
Cr	&	5.62	&$	-0.21	$&	0.13	&$	-0.35	$&	0.13	&$	-0.53	$&	0.08	&$	-0.45	$&	0.29	&$	-0.79	$&	0.07	&$	-0.37	$&	0.15	&$	-0.48	$&	0.08	&$	-0.51	$&	0.11	&$	-0.50	$&	0.12	&$	-0.34	$&	0.11	&$	-0.43	$&	0.13	\\
Mn	&	5.42	&$	-0.25	$&	0.23	&$	-	$&	$-$	&$	-1.16	$&	0.30	&$	-0.59	$&	0.31	&$	-0.73	$&	0.31	&$	-	$&	$-$	&$	-0.97	$&	0.12	&$	-0.45	$&	0.30	&$	-0.70	$&	0.26	&$	-0.57	$&	0.12	&$	-0.03	$&	0.18	\\
Co	&	4.94	&$	0.07	$&	0.08	&$	-	$&	$-$	&$	0.30	$&	0.31	&$	-	$&	$-$	&$	-	$&	$-$	&$	-	$&	$-$	&$	0.15	$&	0.15	&$	0.06	$&	0.07	&$	0.10	$&	0.10	&$	0.09	$&	0.10	&$	0.10	$&	0.11	\\
Ni	&	6.20	&$	0.19	$&	0.09	&$	0.05	$&	0.13	&$	-	$&	$-$	&$	-0.09	$&	0.21	&$	0.13	$&	0.21	&$	0.24	$&	0.12	&$	-0.03	$&	0.11	&$	-0.16	$&	0.13	&$	0.03	$&	0.12	&$	0.11	$&	0.12	&$	-0.07	$&	0.28	\\
Zn	&	4.56	&$	0.83	$&	0.30	&$	0.44	$&	0.27	&$	-	$&	$-$	&$	-	$&	$-$	&$	-	$&	$-$	&$	0.40	$&	0.15	&$	-	$&	$-$	&$	0.29	$&	0.23	&$	0.12	$&	0.25	&$	0.71	$&	0.30	&$	0.43	$&	0.25	\\
Sr	&	2.83	&$	-0.73	$&	0.17	&$	-	$&	$-$	&$	<-2.07	$&	$-$	&$	<-1.76	$&	$-$	&$	-1.25	$&	0.30	&$	-	$&	$-$	&$	-1.12	$&	0.19	&$	-0.58	$&	0.22	&$	-0.06	$&	0.21	&$	-1.22	$&	0.19	&$	-0.28	$&	0.26	\\
Ba	&	2.27	&$	-1.12	$&	0.15	&$	-0.46	$&	0.13	&$	<-1.21	$&	$-$	&$	<-1.12	$&	$-$	&$	-0.62	$&	0.14	&$	<-1.41	$&	$-$	&$	-1.12	$&	0.09	&$	<-1.67	$&	$-$	&$	-0.60	$&	0.10	&$	-0.96	$&	0.15	&$	<-1.38	$&	$-$	\\
\hline

\multicolumn{24}{l}{LR sample}\\																	
\hline			
	&		&	\multicolumn{2}{c}{DR0080}			&	\multicolumn{2}{c}{ES03170}			&&&&&&&&&&&&&&&&&&\\
El.	&	$\log\epsilon_\odot$	&	[X/Fe]	&	$\delta_\text{[X/Fe]}$	&	[X/Fe]	&	$\delta_\text{[X/Fe]}$	&&&&&&&&&&&&&&&&&&\\
\hline											
Fe$^a$	&	7.46	&$	-3.00	$&	0.14	&$	-3.00	$&	0.14	&&&&&&&&&&&&&&&&&&\\
C$_{obs}$	&	8.46	&$	0.49	$&	0.17	&$	-0.46	$&	0.17	&&&&&&&&&&&&&&&&&&\\
C$_{corr}$	&	8.46	&$	1.02	$&	0.17	&$	-0.12	$&	0.17	&&&&&&&&&&&&&&&&&&\\
Ba          &   2.27    &    $<-0.5$        &   -       &      -     & - &&&&&&&&&&&&&&&&&&\\
\hline\hline																							\multicolumn{24}{l}{\fontsize{6.1}{4}\selectfont $^{a)}$ [Fe/H] is listed instead of [X/Fe].}\\

\end{tabular}

\end{sidewaystable*}

%% file: tables/AS0039.tex
\begin{table}
\caption{The NLTE chemical abundances of the star AS0039.
}
\label{tab:as0039}
\normalsize
\tabcolsep=0.08cm
\renewcommand{\arraystretch}{1.2}

\centering
\begin{tabular}{lcccc}
\hline\hline
El.	&	$\log\epsilon_\odot$	&	$\Delta$[X/H]$_\text{NLTE}$	&	[X/Fe]$_\text{NLTE}$	&	$\delta_\text{[X/Fe]}$	\\
\hline
Fe$^a$	& 7.46  &$	+0.26	$&$	-3.64	$&	0.11	\\
C$_\textsl{corr}$	&	8.46&$	-	    $&$	-0.59	$&	0.20	\\
Na	&	6.22	&$	+0.02	$&$	-0.53	$&	0.07	\\
Mg	&	7.55	&$	+0.19	$&$	+0.15	$&	0.05	\\
Al	&	6.43	&$	+1.00	$&$	+0.05	$&	0.10	\\
Si	&	7.51	&$	-0.16	$&$	-0.35	$&	0.15	\\
Ca	&	6.30	&$	+0.28	$&$	+0.08	$&	0.11	\\
Sc	&	3.14	&$	-	    $&$	-0.36	$&	0.08	\\
Ti	&	4.97	&$	+0.03	$&$	+0.40	$&	0.06	\\
Cr	&	5.62	&$	+0.76	$&$	+0.29	$&	0.13	\\
Mn	&	5.42	&$	+0.47	$&$	-0.04	$&	0.23	\\
Co	&	4.94	&$	+0.78	$&$	+0.59	$&	0.08	\\
Ni	&	6.20	&$	+0.26	$&$	+0.19	$&	0.09	\\
Zn	&	4.56	&$	+0.06	$&$	+0.63	$&	0.30	\\
\hline \hline 
\multicolumn{5}{l}{\fontsize{6.1}{4}\selectfont $^{a)}$ [Fe/H] is listed instead of [X/Fe].}\\

\end{tabular}

\end{table}